\documentclass[conference]{IEEEtran}
\IEEEoverridecommandlockouts

\usepackage{cite}
\usepackage{amsmath,amssymb,amsfonts}
\usepackage{graphicx}
\usepackage{textcomp}
\usepackage{xcolor}

\PassOptionsToPackage{hyphens}{url}

\usepackage{subcaption}
\usepackage{fancyhdr}
\usepackage[colorlinks=true, citecolor=purple, linkcolor=violet, urlcolor=purple]{hyperref}
\usepackage{csquotes}
\usepackage{tikz}
\usetikzlibrary{positioning, arrows.meta, fit, backgrounds}
\usepackage[switch]{lineno}
\usepackage{listings}
\usepackage{soul}
\usepackage{booktabs}
\usepackage{algorithm}
\usepackage{algpseudocode}
\usepackage[normalem]{ulem}
\usepackage{comment}
\usepackage{cuted}
\usepackage{tablefootnote}
\usepackage[table]{xcolor}
\usepackage{tabularx}
\usepackage{array}
\usepackage{multirow}
\usepackage{pifont}
\usepackage{makecell}
\usepackage{threeparttable}

\newcommand{\Arcalis}{\textit{Arcal\'{\i}s}}

\newcommand{\cmark}{\textcolor{green!60!black}{\ding{51}}} 
\newcommand{\xmark}{\textcolor{red!70!black}{\ding{55}}}   

\newcommand{\SoftwareSym}{\color{teal}$\mathsf{S}$}
\newcommand{\NICSym}{\color{violet}$\mathsf{N}$}
\newcommand{\NearLLCSym}{\color{blue}$\mathsf{NL}$}
\newcommand{\FPGASym}{\color{purple}$\mathsf{F}$}
\newcommand{\OnChipSym}{\color{black}$\mathsf{OC}$}

\definecolor{codeblue}{RGB}{0,20,255}
\definecolor{codegreen}{RGB}{0,128,0}
\definecolor{codegray}{RGB}{128,128,128}
\definecolor{codepurple}{RGB}{128,0,128}
\definecolor{codeorange}{RGB}{200,50,0}
\definecolor{backcolour}{RGB}{252,252,252}

\lstdefinestyle{arcalisstyle}{
    backgroundcolor=\color{backcolour},
    commentstyle=\color{codegreen}\itshape,
    keywordstyle=\color{codeblue},
    numberstyle=\tiny\color{codegray},
    stringstyle=\color{codepurple},
    identifierstyle=\color{black},
    basicstyle=\ttfamily\scriptsize,
    breakatwhitespace=false,
    breaklines=true,
    captionpos=b,
    keepspaces=true,
    numbers=left,
    numbersep=8pt,
    showspaces=false,
    showstringspaces=false,
    showtabs=false,
    tabsize=4,
    frame=None, 
    frameround=tttt,
    rulecolor=\color{codegray!40},
    aboveskip=12pt,
    belowskip=12pt,
    xleftmargin=10pt,
    xrightmargin=5pt,
    framexleftmargin=8pt,
    morekeywords={bool, void, int, for, if, break, return, while, else, struct, unsigned},
    emph={appCallEngineRead, netCallEngineRead, appCallEngineWrite, netCallEngineWrite}, 
    emphstyle=\color{codeorange}
}

\definecolor{commentgreen}{rgb}{0,0.6,0}

\newcommand{\rev}[1]{{#1}}
\newcommand\revB[1]{\noindent{{#1}}}
\newcommand\revPM[1]{\noindent{{#1}}}
\newcommand{\del}[1]{}

\definecolor{gray}{RGB}{217, 217, 217}
\newcommand{\graycircle}[1]{%
  \tikz[baseline=(char.base)]{
    \node[
      shape=circle,
      draw,
      fill=gray,
      inner sep=1pt,
      outer sep=0pt
    ] (char) {#1};
  }%
}

\newcommand{\whitecircle}[1]{%
  \tikz[baseline=(char.base)]{
    \node[
      shape=circle,
      draw,
      fill=white,
      inner sep=1pt,
      outer sep=0pt
    ] (char) {#1};
  }%
}

\newcommand{\darkcircle}[1]{%
  \tikz[baseline=(char.base)]{
    \node[
      shape=circle,
      draw,
      fill=black,
      text=white,
      inner sep=1pt,
      outer sep=0pt
    ] (char) {#1};
  }%
}

\newcommand{\bluecircle}[1]{%
  \tikz[baseline=(char.base)]{
    \node[
      shape=circle,
      draw=black,
      fill={rgb:red,0;green,90;blue,192},
      text=white,
      inner sep=1pt,
      outer sep=0pt
    ] (char) {#1};
  }%
}

\newcommand{\winecircle}[1]{%
  \tikz[baseline=(char.base)]{
    \node[
      shape=circle,
      draw=black,
      fill={rgb:red,180;green,0;blue,30},
      text=white,
      inner sep=1pt,
      outer sep=0pt
    ] (char) {#1};
  }%
}

\newcommand{\postservice}{PostStorage}
\newcommand{\memcservice}{Memcached}
\newcommand{\uidservice}{UniqueId}
\newcommand{\urlservice}{UrlShorten}
\newcommand{\utservice}{UserTimeline}
\newcommand{\uc}{\texttt{UC}}

\def\BibTeX{{\rm B\kern-.05em{\sc i\kern-.025em b}\kern-.08em
    T\kern-.1667em\lower.7ex\hbox{E}\kern-.125emX}}
\begin{document}

\title{\textbf{Arcalís}: \textbf{A}ccelerating \textbf{R}emote Procedure \textbf{Ca}lls Using a \textbf{Lí}ghtweight Near-Cache \textbf{S}olution\\
}

\author{\IEEEauthorblockN{Johnson Umeike\textsuperscript{*}}\thanks{* Part of this work was done during an internship at AMD.} 
\thanks{$\dagger$ The opinions expressed are the author's alone and do not represent AMD's \mbox{\quad\;\,} views or indicate any future AMD software or hardware policy.}
\IEEEauthorblockA{\textit{Department of Computer Science} \\
\textit{University of Maryland (UMD)}\\
College Park, MD, USA \\
jumeike@umd.edu}
\and
\IEEEauthorblockN{Pongstorn Maidee\textsuperscript{$\dagger$}}
\IEEEauthorblockA{\textit{Research and Advanced Development (RAD)} \\
\textit{Advanced Micro Devices, Inc. (AMD)}\\
San Jose, CA, USA \\
pongstorn.maidee@amd.com}
\and
\IEEEauthorblockN{Bahar Asgari}
\IEEEauthorblockA{\textit{Department of Computer Science} \\
\textit{University of Maryland (UMD)}\\
College Park, MD, USA \\
bahar@umd.edu}
}

\maketitle


\begin{abstract}

\textit{Modern microservices increasingly depend on high-performance remote procedure calls (RPCs) to coordinate fine-grained, distributed computation. As network bandwidths continue to scale, the CPU overhead associated with RPC processing, particularly serialization, deserialization, and protocol handling, has become a critical bottleneck. This challenge is exacerbated by fast user-space networking stacks such as DPDK, which expose RPC processing as the dominant performance limiter. 
While prior hardware accelerators have explored NIC-attached and FPGA-based offloads, these approaches remain \revB{farther from the cache hierarchy, so the frequent data accesses during RPC processing each pay an extra interconnect traversal cost that inflates RPC time.} Therefore, RPC handling should occur as close as possible to the cache; however, a near-cache solution must be small, hence practical and deployable. Our key insight to enable such a solution is to take advantage of a reconfigurable accelerator that can be configured specifically for the services currently running on the CPUs. 
We present \Arcalis{}, a near-cache RPC accelerator that positions a lightweight hardware engine adjacent to the last-level cache (LLC). \Arcalis{} offloads RPC processing to dedicated microengines that operate with cache-line latency while preserving programmability. By decoupling RPC processing logic, enabling microservice-specific execution, and positioning itself near the LLC, \Arcalis{} achieves a 1.72-4.91$\times$ end-to-end speedup compared to the CPU baseline, significantly reduces microarchitectural overhead by up to 88\%, and achieves up to a 1.67$\times$ higher throughput than prior solutions. These results highlight the potential of near-cache RPC acceleration as a practical solution for high-performance microservice deployment.}

\end{abstract}

\begin{IEEEkeywords}
Datacenter, hardware accelerator, hardware-software codesign, microservices, near-cache processing, programmable logic, reconfiguration, remote procedure call.
\end{IEEEkeywords}

\section{Introduction}

\label{sec:introduction}

\begin{figure}[t]
    \centering
    \includegraphics[width=0.9\linewidth]{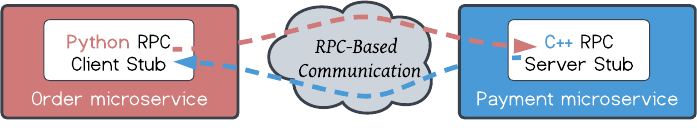}
    \caption{\textit{RPC enables cross-language, platform-independent communication between heterogeneous microservices.}}
    \label{fig:microservic-overview}
    \vspace{-1em}
\end{figure}

Modern datacenter applications are increasingly built using \textit{microservices}, which are independently deployable components that communicate through remote procedure calls (RPCs). Compared to monolithic architectures, microservices offer better modularity, scalability, and deployment flexibility~\cite{gan-asplos19-deathstarbench},\cite{sriraman-iiswc18-microsuite}. Figure~\ref{fig:microservic-overview} illustrates how microservices interact through language-agnostic RPC stubs, enabling a Python-based client to invoke functionality implemented in a C++ service. This cross-language abstraction simplifies distributed development but introduces substantial software and communication overhead into each request, which is now commonly referred to as \textit{``RPC tax''}~\cite{pourhabibi-micor21-cerebros, kanev-isca15-profilingwsc, seemakhupt-sosp23-profiling}. The RPC tax manifests in multiple forms across the system stack. At the application level, microservices spend 40-90\% of their execution time in RPC processing rather than in business logic~\cite{pourhabibi-micor21-cerebros}. At the microarchitectural level, RPC workloads exhibit poor instruction locality, high branch misprediction rates, and memory-bound execution patterns~\cite{sriraman-iiswc18-microsuite, mirhosseini2019enhancing}.

The problem is exacerbated by the growing deployment of high-performance networking technologies. Userspace networking frameworks such as DPDK~\cite{dpdk}, Snap~\cite{marty2019snap}, IX~\cite{belay2014ix}, ZygOS~\cite{prekas2017zygos}, and hardware-accelerated transports such as RDMA~\cite{nvidia_rdma_routed}, InfiniBand~\cite{nvidia_infiniband}, RoCE~\cite{nvidia_roce} eliminate kernel bottlenecks but expose RPC processing as the dominant performance limiter~\cite{lazarev-asplos21-dagger, zhang-hpca25-rpcnic}. The serialization/deserialization layer alone consumes up to 94\% of RPC processing cycles~\cite{pourhabibi-micor21-cerebros}. Our microarchitectural analysis (\S\ref{sec:motivation}) reveals that even with optimized userspace networking stacks like DPDK~\cite{dpdk}, up to 67.6\% of pipeline slots are lost to frontend bottlenecks and memory hierarchy stalls, with only 47.9\% achieving productive instruction retirement in the best case. This creates a fundamental mismatch between network capability and application performance.
To address these limitations, researchers have proposed various RPC acceleration approaches, each with distinct trade-offs (summarized in Table~\ref{tab:rpc_table}).

\noindent\textbf{NIC-attached accelerators} such as Optimus Prime~\cite{pourhabibi2020optimus}, Cerebros~\cite{pourhabibi-micor21-cerebros}, and NeBuLa~\cite{sutherland-isca20-nebula} \revB{integrate} RPC processing \revB{with on-chip network interfaces~\cite{novakovic2014sonuma}}, significantly reducing the CPU overhead. \revB{However, while these designs are schema-programmable and adapt to new field layouts in software, their datapath is fixed in silicon, so supporting a new or rapidly evolving RPC framework eventually requires a hardware change that fixed silicon cannot accommodate without a chip redesign cycle, which could take years~\cite{zhang-hpca25-rpcnic}.} 

\noindent\textbf{On-chip accelerators} such as ProtoAcc~\cite{karandikar-micro21-protoacc} couple custom serialization and deserialization logic to the processor core via a co-processor interface, requiring custom ISA extensions and tight per-core integration. This creates deployment barriers, as applications must be recompiled 
to use the accelerator. 

\noindent\textbf{FPGA-based solutions} provide reconfigurability and high throughput. Dagger~\cite{lazarev-asplos21-dagger} leverages specialized NUMA interconnects (UPI) to achieve 400\,ns access latency; however, its reliance on a non-commodity interconnect limits deployability and complicates integration in standard datacenter servers. \revB{Observing that RPC frameworks evolve faster than fixed-silicon accelerators can be updated,} RpcNIC~\cite{zhang-hpca25-rpcnic} proposes a more general, \revB{reconfigurable}, but slower PCIe-connected architecture ($\sim$900ns interconnect latency~\cite{neugebauer-sigcomm18-pcieperf}), but its dependence on CPU-integrated memcpy engines necessitates splitting RPC processing between the host CPU and the accelerator, preventing complete offload. 

\noindent\textbf{Software-only approaches} such as FARM~\cite{dragojevic2014farm}, eRPC~\cite{kalia2019datacenter} and FaSST~\cite{kalia2016fasst} optimize RPC performance through kernel-bypass networking and efficient software stacks. While offering low deployment barriers, these solutions still consume substantial CPU resources and cannot match the performance isolation of hardware-assisted designs under high loads.

\begin{table}[t]
\centering
\begin{threeparttable}
\caption{\textit{Comparison of RPC acceleration proposals.}}
\label{tab:rpc_table}
\renewcommand{\arraystretch}{1.05}
\setlength{\tabcolsep}{2.5pt}
\scriptsize
\begin{tabularx}{\columnwidth}{l|>{\centering\arraybackslash}X|c|c|c}
\hline
\thead{\textit{RPC Acceleration}\\\textit{Proposal}} &
\thead{\textit{Design}\\\textit{Features}} &
\thead{\textit{\revB{Reconfigurable}}\\\textit{\revB{Datapath}}} &
\thead{\textit{Full RPC}\\\textit{Offload}} &
\thead{\textit{Cache-Line}\\\textit{Access Latency}} \\
\hline \hline
eRPC~\cite{kalia2019datacenter}             & $\textbf{\SoftwareSym}$                                  & N/A & N/A   & N/A   \\ \hline
Optimus Prime~\cite{pourhabibi2020optimus}  & $\textbf{\revB{\OnChipSym}},\textbf{\NICSym}$                                       & \xmark & \xmark & \xmark \\ \hline
Cerebros~\cite{pourhabibi-micor21-cerebros} & $\textbf{\OnChipSym},\textbf{\NICSym}$                                       & \xmark & \cmark & \xmark \\ \hline
ProtoAcc~\cite{karandikar-micro21-protoacc} & $\textbf{\OnChipSym}$                                    & \xmark & \xmark & \cmark \\ \hline
Dagger~\cite{lazarev-asplos21-dagger}       & $\textbf{\FPGASym}$                                      & \cmark & \cmark & \xmark \\ \hline
RpcNIC~\cite{zhang-hpca25-rpcnic}           & $\textbf{\NICSym}, \textbf{\FPGASym}$                    & \cmark & \cmark & \xmark \\ \hline
NeBuLa~\cite{sutherland-isca20-nebula}      & $\textbf{\OnChipSym},\textbf{\NICSym}$                                       & \xmark & \cmark & \xmark \\ \hline \hline
\textbf{\Arcalis}                           & $\textbf{\OnChipSym}, \textbf{\FPGASym}, \textbf{\NearLLCSym}$ & \cmark & \cmark & \cmark \\ \hline \hline
\end{tabularx}
\begin{tablenotes}[flushleft]
\scriptsize
\item \textbf{\SoftwareSym} = \textit{Software-based},
\textbf{\NICSym} = \textit{NIC-attached},
\textbf{\FPGASym} = \textit{FPGA-based},
\textbf{\OnChipSym} = \textit{On-chip},
\textbf{\NearLLCSym} = \textit{Near-LLC}.
\item \end{tablenotes}
\end{threeparttable}
\vspace{-2em}
\end{table}


The fundamental unsolved challenge is that existing solutions either sacrifice \revB{adaptability} for performance (i.e., fixed-\revB{silicon} accelerators), incur data movement overhead, or sacrifice performance for deployability (i.e., software approaches). Moreover, partially offloading (de)serialization alone is insufficient because the RPC logic interleaved with business logic on the CPU disrupts instruction and data locality, amplifying memory-bound behavior regardless of how much the (de)serialization overhead is removed (see \S\ref{sec:motivation}). 

Recent I/O techniques allow NICs to place data directly into the CPU cache hierarchy. To fully eliminate unnecessary data movement, RPC handling should occur as close to the cache as possible. The main obstacle to achieving this efficiency, however, is that a generic solution is too large to be integrated at that proximity. Running a service on servers that recently executed it improves performance by avoiding cold starts\mbox{~\cite{2025ColdStarts}}. In particular, stateful microservices—databases and key-value stores such as Memcached—are often deployed on dedicated servers to avoid migrating large amounts of state\mbox{~\cite{Luo2021AlibabaMicroservice}}. In those cases, it is sensible to tailor the accelerator specifically for the services currently running on the CPUs.

To solve the unsolved challenge of RPC acceleration—optimizing \revB{adaptability} and data movement without sacrificing either one—and to efficiently enable a RPC accelerator deployable near the cache, our key insight is to take advantage of reconfigurable computing to configure an accelerator specifically for the services currently running on the CPUs. To achieve this goal, we propose \Arcalis{\footnote{\Arcalis{} is a subgiant star in the constellation Boötes.}. To eliminate data movement overhead, \Arcalis{} is a lightweight hardware engine located adjacent to the last-level cache, featuring reconfigurable \texttt{RxEngine}/\texttt{TxEngine} processing pipelines that achieve cache-line access latency while maintaining programmability through systematic hardware-software co-design. \revB{Although prior on-chip NIC-attached accelerators are also integrated on-die, they are placed at the chip's edge~\mbox{\cite{sutherland-isca20-nebula, pourhabibi-micor21-cerebros, pourhabibi2020optimus}}, whereas \Arcalis{} sits beside the LLC slices, so it reaches cache-resident data over fewer on-chip hops and at lower access latency.} 
To design \Arcalis{} efficiently, we perform a comprehensive analysis of RPC bottlenecks, revealing fundamental microarchitectural inefficiencies in modern processor pipelines and demonstrating that near-cache acceleration fundamentally improves processor efficiency, reducing instruction count by \rev{23}--86\%, load-to-use latency by \rev{22}--88\%, and cache access overhead by \rev{20}--87\%. Evaluation shows that \Arcalis{} achieves 1.72--4.\rev{91}$\times$ end-to-end performance improvements.

\section{Background}\label{background}

\begin{figure}[t]
    \centering
    \includegraphics[width=1\linewidth]{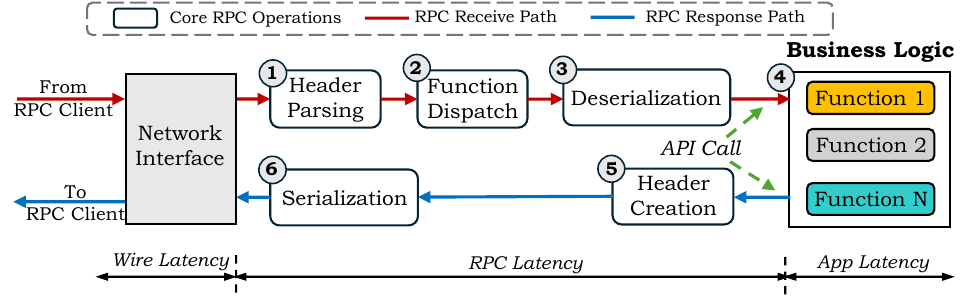}
    \caption{\textit{RPC processing pipeline in modern microservices deployments. Stages 1-3, 5-6 are implemented within standard RPC frameworks~\cite{thrift,grpc}.}}
    \label{fig:rpc-pipeline}
    \vspace{-1em}
\end{figure}

This section outlines the architectural shifts and system level trends that shape the modern RPC execution path. Our goal is to expose structural inefficiencies that arise from treating communication and computation as separate layers, inefficiencies that become unsustainable as datacenter networks outpace memory systems and software stacks. These trends highlight the need to revisit long standing assumptions about how performance critical communication workloads interact with hardware.

\vspace{0.2em}
\noindent\textbf{RPC Processing in Modern Microservices.}
Early RPC frameworks such as Sun RPC~\cite{sunrpc_wiki} and CORBA~\cite{corba_wiki} prioritized simplicity and local area network (LAN) compatibility, using fixed-format messages and basic marshalling schemes tailored for homogeneous systems. In contrast, modern cloud applications rely on microservices, where high-frequency RPCs orchestrate fine-grained interactions among loosely coupled services~\cite{zhou2018benchmarking, zhang2019muqsim, wang-acm22-dcaperf, gan-asplos19-deathstarbench}. These RPCs operate with tight latency budgets~\cite{zhao2022altocumulus} and are mediated by frameworks such as Thrift~\cite{thrift}, gRPC~\cite{grpc}, and Cap’n Proto~\cite{capnproto2025}, which support schema evolution, type safety, and multi-language bindings. However, these features introduce layered processing overheads that translate into real resource overheads.
As Figure~\ref{fig:rpc-pipeline} shows, a typical RPC call at the server traverses six structured software stages: header parsing (\graycircle{1}) extracts metadata such as function name and request ID; function dispatch (\graycircle{2}) invokes the appropriate handler from a function table; deserialization (\graycircle{3}) reconstructs in-memory objects from the wire format; business logic (\graycircle{4}) performs the core computation (e.g., SET or GET in the case of memcached); serialization (\graycircle{5}) encodes the response back to wire format; and header creation (\graycircle{6}) attaches protocol metadata for reply transmission. These steps involve predictable but compute- and memory-intensive tasks, including schema-driven field parsing and dynamic function resolution~\cite{khairy2022simr, pourhabibi2020optimus}. Profiling studies show that the core RPC operations account for a significant fraction of per-request latency and CPU cycles in production systems~\cite{sriraman-iiswc18-microsuite, ueda-iiswc16-workloadcharacterization, gan-cal18-archimpl}, making RPC handling a major source of inefficiency at scale.

\begin{figure}[h]
    \centering
    \includegraphics[width=1\linewidth]{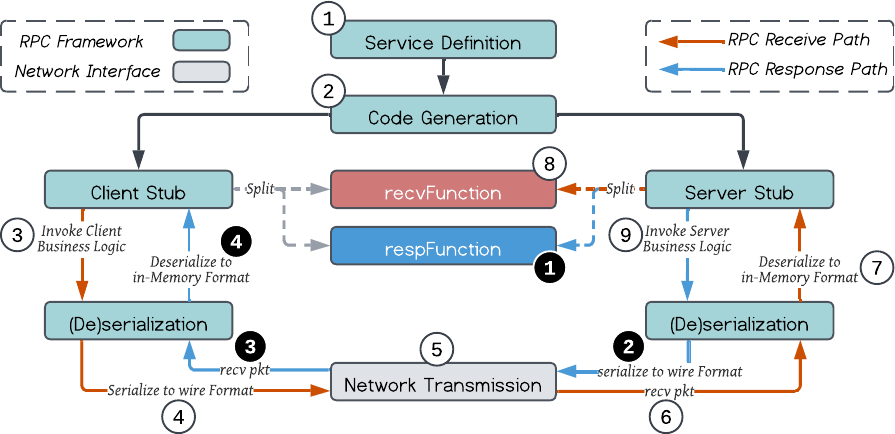}
    \caption{\textit{RPC communication workflow. End-to-end request and response paths through client/server stubs.}}
    \label{fig:rpc-workflow}

\end{figure}

Figure~\ref{fig:rpc-workflow} illustrates the RPC communication workflow. Following service definition and setup (\whitecircle{1}--\whitecircle{2}), the request traverses the receive path: the client stub invokes the RPC (\whitecircle{3}), serializes it to wire format (\whitecircle{4}), and transmits it over the network (\whitecircle{5}). On the server side, the request is received (\whitecircle{6}), its header is deserialized (\whitecircle{7}), and dispatched to the business logic function (\whitecircle{9}) after RPC-side recvFunction execution (\whitecircle{8}). Although the compiler emits a single stub on each side, the execution naturally separates into a \textit{recvFunction}, which handles deserialization of incoming request body and dispatch on the receive path, and a \textit{respFunction}, which handles the serialization of the outgoing response body on the response path. The highlighted dark markers (\darkcircle{1}--\darkcircle{4}) denote the response path, where the response body is serialized at the server, transmitted, and deserialized at the client for application consumption. As we show in \S\ref{sec:operation_and_design}, \Arcalis{} exploits this natural split by mapping \textit{recvFunction} logic to the \texttt{RxEngine} and \textit{respFunction} logic to the \texttt{TxEngine}, enabling complete hardware offload of both paths.

\begin{figure}[t]
\centering
\begin{subfigure}[b]{0.40\columnwidth}
    \centering
    \includegraphics[width=\textwidth]{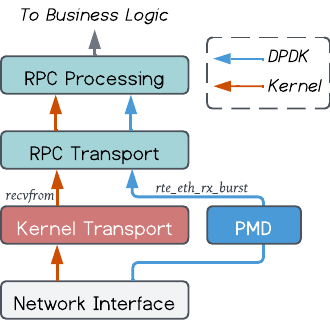}
    \caption{\textit{RPC Dual Paths}}
    \label{fig:rpc_dual_path}
\end{subfigure}
\hfill
\begin{subfigure}[b]{0.58\columnwidth}
    \centering
    \includegraphics[width=\textwidth]{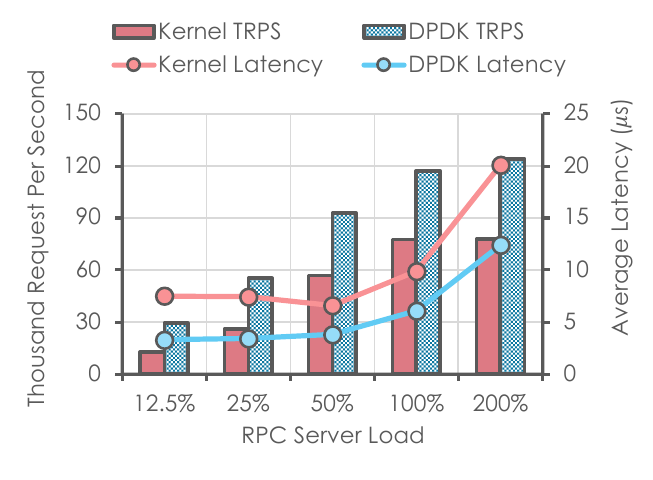}
    \vspace{-2.2em}
    \caption{\textit{Throughput and Average Latency}}
    \label{fig:trps_latency}
\end{subfigure}
\vspace{-0.8em}
\caption{\textit{Effectiveness of kernel-bypass networking for RPC execution: (a) Kernel and DPDK receive RPC processing paths. DPDK directly access the network interface hardware using its poll mode driver (PMD) (b) Throughput (thousand requests per second) and average latency for Memcached RPC (0.8 SET ratio) using DPDK and kernel networking.}}
\vspace{-1em}
\label{fig:userpace_vs_kernel}
\end{figure}

\vspace{0.2em}
\noindent\textbf{Datacenter I/O Optimizations.}
Modern datacenters have adopted high-speed interconnects and kernel-bypass techniques to reduce communication overhead. Technologies such as DPDK~\cite{dpdk}, Snap~\cite{marty2019snap}, and mTCP~\cite{jeong2014mtcp} enable applications to send and receive packets directly from userspace, bypassing the kernel networking stack. These frameworks reduce context switches, memory copies, and interrupt overhead, achieving packet latencies on the order of microseconds and throughput scaling into tens of millions of requests per second~\cite{prekas2017zygos, belay2014ix, ousterhout2019shenango, fried2020caladan}. 

Figure~\ref{fig:rpc_dual_path} differentiates the kernel and DPDK RPC receive paths. In the kernel stack, packets traverse the NIC driver, kernel transport, and socket layer before reaching the RPC runtime. DPDK bypasses these layers entirely, allowing a poll-mode driver (PMD) to pull packets directly into userspace and hand them to the RPC framework, completely avoiding interrupts, data copying, and dynamic \texttt{skb} allocation overheads~\cite{cai-sigcomm21-understandinghoststackoverhead}. Figure~\ref{fig:trps_latency} shows that DPDK achieves higher request throughput and lower average latency across a wide range of server loads. While both paths eventually saturate the core, the kernel stack incurs additional queuing and scheduling overheads, causing latency to rise sharply under load. RDMA-based solutions such as InfiniBand~\cite{nvidia_infiniband} and RoCE~\cite{nvidia_roce} also bypass the kernel network stack, providing direct memory access between machines without CPU intervention. They both leverage specialized NIC hardware and transport engines to perform reliable communication with sub-µs latency.

To meet the demands of high-speed networks, recent innovations have focused on reducing CPU-side memory access costs during packet processing. Techniques such as AMD’s smart data cache injection (SDCI)\footnote{AMD, the AMD Arrow logo and combinations thereof are trademarks of Advanced Micro Devices, Inc.} 
~\cite{amd_sdci}, Intel’s data direct I/O (DDIO)~\cite{intel_ddio}, and Arm’s cache stashing~\cite{arm_cache_stashing} allow network devices to inject incoming packets directly into the processor’s cache. These direct cache access (DCA) mechanisms bypass DRAM, reducing latency and memory bandwidth usage. Inbound DCA writes use a write-allocate/write-update (WAU) policy to place I/O data directly into the LLC~\cite{farshin2020reexamining}. If the cache line already exists, it is updated in place and marked as dirty. Otherwise, it is allocated in a reserved LLC portion. The line remains in cache and is written back to DRAM only upon eviction. By avoiding costly direct memory access (DMA) round-trips to system memory, DCA enhances the efficiency of CPU-bound microservices and enables better overlap between network and computation tasks~\cite{li2015architecting}.

\vspace{0.2em}
\noindent\textbf{Accelerator Integration Models. }The trend toward domain-specific acceleration has driven the exploration of different accelerator placement strategies within the system hierarchy, each offering distinct trade-offs between performance, cost, and integration complexity. These range from peripheral accelerators connected via PCIe~\cite{firestone2018azure, eran2019nica}, to near-memory processing elements positioned at memory controllers~\cite{alian2019netdimm, ghose2019processing}, to on-chip accelerators integrated within the processor package~\cite{caulfield2016cloud, esmaeilzadeh2011dark}. Each placement offers different latency, bandwidth, and coherency characteristics that impact workload suitability. 

For RPC processing specifically, existing approaches span this placement spectrum~\cite{pourhabibi-micor21-cerebros, lazarev-asplos21-dagger, zhang-hpca25-rpcnic, sutherland-isca20-nebula}, but each faces a placement-induced limitation. \revB{Peripheral accelerators traverse an off-die link on every access, whether over PCIe or a NUMA memory interconnect, and pay its latency per RPC field. On-chip designs avoid the off-die link but integrate the accelerator at the on-chip NIC, at the chip edge\mbox{\cite{novakovic2014sonuma, daglis2019rpcvalet, daglis2015manycore}}, leaving it several interconnect hops from the LLC where RPC data is accessed. None places the accelerator adjacent to the LLC, where cache-resident data can be reached at cache-line~latency.}

\section{Opportunities for Near-Cache Acceleration}\label{sec:motivation}
\subsection{Characterizing RPC Processing Bottlenecks}
\begin{figure}[t]
\centering
\begin{subfigure}[b]{0.48\columnwidth}
    \centering
    \includegraphics[width=\textwidth]{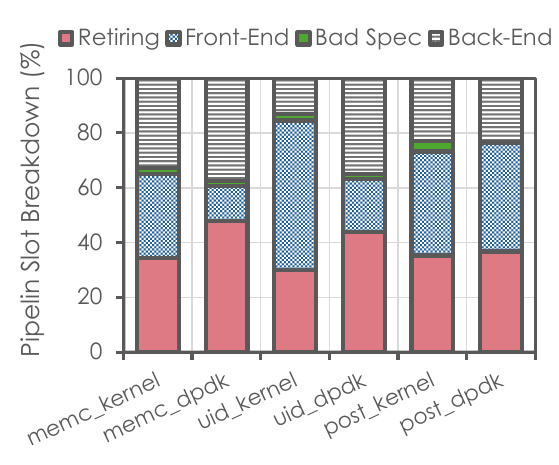}
    \vspace{-1.8em}
    \caption{\textit{Pipeline slot breakdown.}}
    \label{fig:vtune_profile}
\end{subfigure}
\hfill
\begin{subfigure}[b]{0.48\columnwidth}
    \centering
    \includegraphics[width=\textwidth]{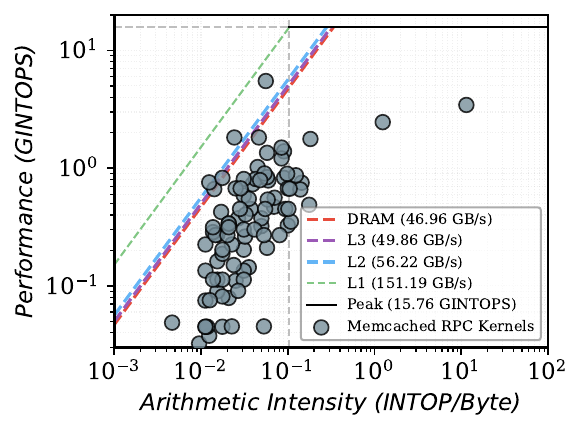}
    \vspace{-1.8em}
    \caption{\textit{MemcachedService Roofline}}
    \label{fig:memc_roofline}
\end{subfigure}
\hfill
\begin{subfigure}[b]{0.48\columnwidth}
    \centering
    \includegraphics[width=\textwidth]{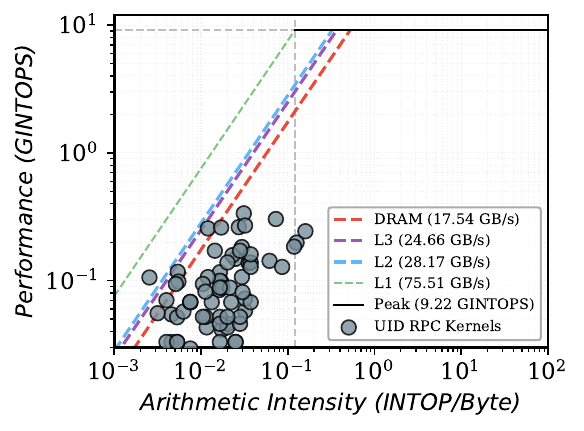}
    \vspace{-1.8em}
    \caption{\textit{UIDService Roofline}}
    \label{fig:uid_roofline}
\end{subfigure}
\hfill
\begin{subfigure}[b]{0.48\columnwidth}
    \centering
    \includegraphics[width=\textwidth]{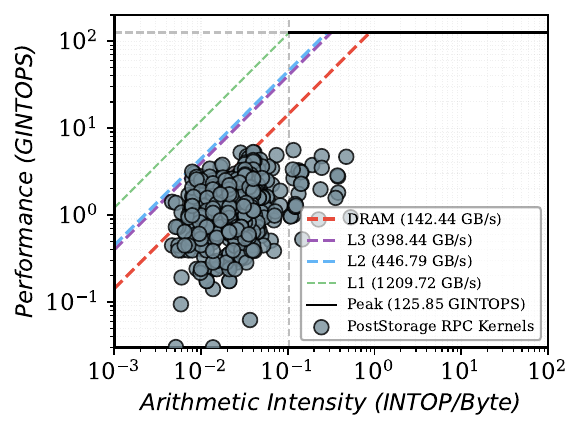}
    \vspace{-1.8em}
    \caption{\textit{PostStorageService Roofline}}
    \label{fig:post_roofline}
\end{subfigure}
\caption{\textit{Performance characterization of RPC microservices: (a) Pipeline slot utilization showing front-end and back-end stalls, (b)-(d) Roofline analysis showing memory-boundedness.}}
\label{fig:Perf_characterization}
\vspace{-1em}
\end{figure}

CPUs are inefficient at processing RPCs due to microarchitectural mismatches between their general-purpose design and the characteristics of RPC workloads. Prior profiling studies of WSC systems reveal that RPC execution suffers from high instruction-fetch overheads, frontend stalls, cache misses, and backend latency~\cite{seemakhupt-sosp23-profiling, gan-cal18-archimpl, shuang-iiswc17-bigsmall, pourhabibi-micor21-cerebros}. Over 30\% of pipeline slots are often stalled at the frontend, and backend cycles are frequently consumed by L1 cache misses and TLB pressure~\cite{kanev-isca15-profilingwsc}. These issues are amplified in microservice-style designs with frequent inter-service RPCs. To understand the microarchitectural inefficiencies of RPC execution, we profile $\memcservice$~\cite{memcached} and representative microservices from DeathStarBench~\cite{gan-asplos19-deathstarbench}, including $\postservice$ and $\uidservice$, on a server hosted on a research testbed~\cite{Duplyakin+:ATC19}, whose configuration is shown in Table~\ref{tab:server_config}.

\begin{table}[h]
\centering
\caption{\textit{Server hardware configuration used for evaluation.}}
\vspace{-0.2em}
\footnotesize
\begin{tabular}{|l|l|}
\toprule
\textbf{Component} & \textbf{Specification} \\ 
\midrule
CPU  & \textit{Intel Xeon Silver 4314 (2.4 GHz, 16 cores / 32 threads)} \\
Memory & \textit{128 GB DDR4--3200 ECC (8 channels, 170.6 GB/s peak)} \\
NIC & \textit{Dual-port Mellanox ConnectX-6 DX (100 GbE per port)} \\
\hline
\end{tabular}
\label{tab:server_config}
\vspace{-0.2em}
\end{table}

Figure~\ref{fig:vtune_profile} shows the pipeline breakdown across six RPC services in our study. In the best case, only 47.9\% of slots retire instructions; the rest are lost to frontend bottlenecks, branch resteers, and backend stalls. Notably, up to 37.6\% of cycles are spent waiting on the memory hierarchy, including DRAM latency, store buffers, and L1-bound stalls. These results persist even when kernel overhead is removed via userspace networking. While prior works examine kernel-mode RPC stacks in isolation, our study is the first to characterize the microarchitectural impact of efficient network stacks such as DPDK on RPC. 

Figures~\ref{fig:memc_roofline}--\ref{fig:post_roofline} show that RPC kernels operate at very low arithmetic intensities, between $10^{-3}$ and $10^{1}$ \rev{giga integer operations per byte} (GINTOP/Byte), placing most of them squarely in the memory-bound region of the roofline. Their achieved performance remains below the compute roof and never approaches the L1 or L2 bandwidth ceilings, indicating that computation is not the limiting factor. Instead, these kernels are dominated by fine-grained metadata accesses and pointer-heavy operations that generate poor spatial locality and frequent cache misses. This \rev{observed} behavior aligns with the pipeline-slot breakdown and highlights that RPC execution on general-purpose cores is fundamentally constrained by memory latency rather than compute throughput.

An additional source of inefficiency arises from the way RPC processing is interleaved with the business logic during execution. This per-request mixing introduces transient instruction and data accesses that disrupt locality, evict useful cache lines, and exacerbate the observed memory-bound behavior. This increases I-cache and ITLB pressure and contributes to memory stalls. Prior acceleration efforts that focus only on (de)serialization overheads~\cite{pourhabibi2020optimus, wolnikowski-hotos21-zerializer} do not eliminate these effects because the CPU still executes the RPC path. Together, these observations point to the need for moving RPC handling closer to where the data lives, thereby avoiding these costly stalls and freeing the core entirely for the business logic.

\subsection{When Zero-Copy Is Not Zero-Cost}
\label{sec:zero-copy-overhead}
High-performance networking frameworks such as DPDK eliminate the kernel from the critical path and leverage zero-copy DMA techniques to boost throughput. Prior work has thoroughly characterized kernel-mode RPC overheads, identifying memory copies and TCP/IP stack processing as dominant bottlenecks on the receiver side~\cite{cai-sigcomm21-understandinghoststackoverhead, skiadopoulos-osdi24-flexiblehostnetworking}. DPDK mitigates these issues using hugepages and direct NIC-to-user memory access, dramatically improving packet delivery. However, as shown in Fig.~\ref{fig:rpc_cpu_breakdown}, removing kernel overheads now exposes the RPC layer as the dominant bottleneck.

\begin{figure}[t]
  \centering
  \includegraphics[width=1\linewidth]{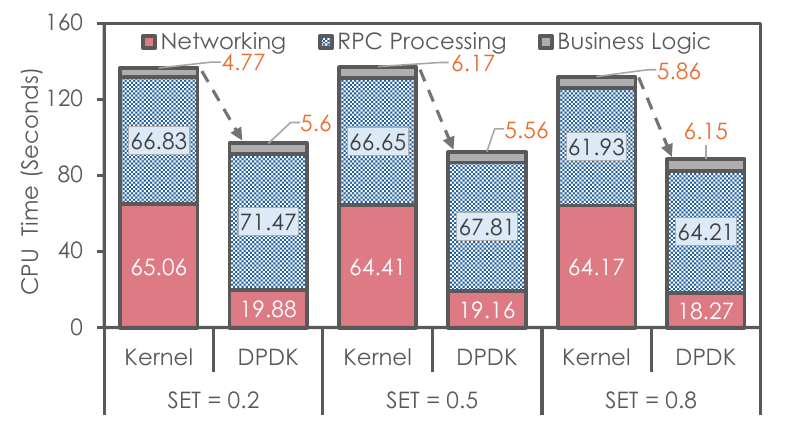}
  \vspace{-1.2em}
  \caption{\textit{CPU time breakdown for Memcached processing 10 million RPC requests across different SET ratios.}} 
  \label{fig:rpc_cpu_breakdown}
  \vspace{-1em}
\end{figure}

We profile CPU time across the network stack, Thrift RPC layer, and business logic while serving 10 million requests in $\memcservice$ microservice under different set ratios (20\%, 50\%, and 80\%). When using kernel networking, a large portion of CPU time ($\sim$50\%) is spent in the slow kernel network stack. With DPDK~\cite{dpdk}, the kernel overhead is largely eliminated, reducing total network time by 3.38$\times$ on average. However, the time spent in RPC processing slightly increases and becomes the dominant component of the total execution time, while the business logic component remains effectively unchanged. This shift highlights that while efficient networking stacks improve packet handling efficiency, it exposes the (de)serialization overheads in RPC, revealing the inefficiency of executing these tasks entirely on general-purpose CPU cores.

Moreover, at higher packet ingress rates, the CPU becomes more exposed to frequent cache line loads and memory pressure. This mirrors the broader “datacenter tax” effect, where even efficient packet delivery translates into heavier application processing burdens~\cite{belay2014ix, ousterhout2019shenango}. Our results empirically reinforce this by showing that, despite lower network layer overhead, efficient networking stacks inflate the total CPU burden for processing RPCs. 

\subsection{Why DMA Isn't enough}
\label{sec:dma-limitation}

DMA-attached accelerators, such as SmartNICs and PCIe-based FPGAs, are a common method for offloading compute and packet processing tasks in datacenters. \rev{These accelerators process data on-device and deliver results to host memory via DMA over high-latency interconnects such as PCIe}. While effective for batch or coarse-grained workloads, this model introduces higher overheads for fine-grained microservices. Because these accelerators are decoupled from the CPU cache hierarchy, each invocation incurs buffer orchestration, memory synchronization, and data transfer latency~\cite{shao-micro16-gem5aladdin}. These costs are particularly pronounced for short-lived RPCs, which involve small messages and frequent core interactions. 

\rev{Unlike bulk inbound packet DMA, RPC deserialization writes individual fields to non-contiguous host memory locations, generating numerous small DMA writes that quickly saturate PCIe transaction rate, partially mitigated in prior work via batched DMA~\mbox{\cite{zhang-hpca25-rpcnic}}. On the transmit path, application objects are scattered across host memory following pointer-dependent references, where each field's address is unknown until the prior field is read, preventing batching and incurring the full interconnect round-trip latency per field~\mbox{\cite{zhang-hpca25-rpcnic}.}}

To address these limitations, many recent efforts across domains, including graph analytics and stencil computation, have proposed near-cache or cache-coherent accelerators~\cite{izmirli-cal25-halis, denzler-ieee23-casper, nori-isca21-reduct, wang-hpca22-nearstream} to integrate compute units closer to the core, maintain coherence with the CPU cache hierarchy, and eliminate data movement setup overhead.

\section{Arcalís: A Near-Cache RPC Accelerator}
\begin{figure*}[t]
  \centering
  \begin{subfigure}[t]{0.67\textwidth}
    \centering
    \includegraphics[width=\linewidth]{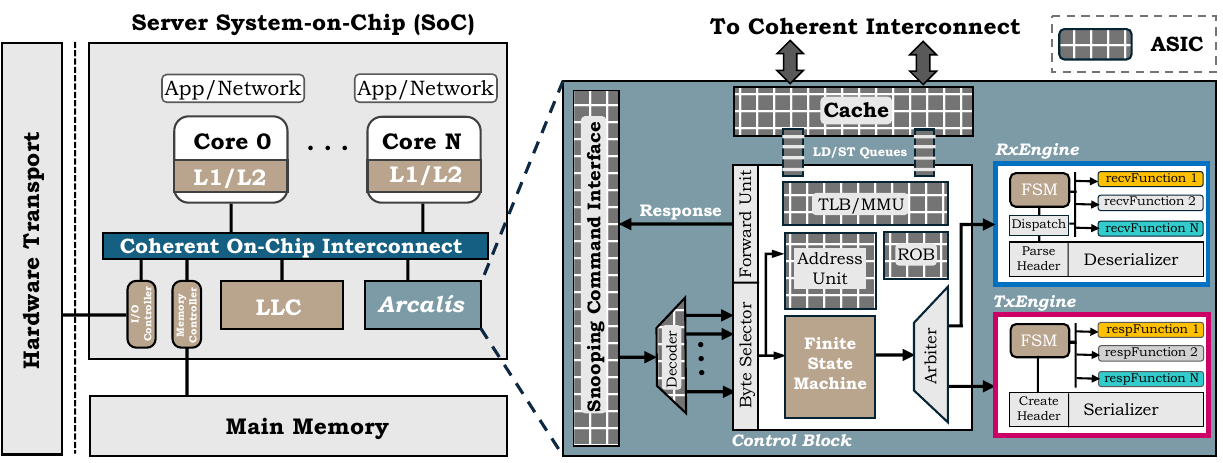}
    \caption{\textit{\Arcalis{}' architecture, showing FLR or ASICs as shaded. Other components are the RLR.}}
    \label{fig:arcalis-architecture}
  \end{subfigure}
  \hfill
  \begin{subfigure}[t]{0.32\textwidth}
    \centering
    \includegraphics[width=\linewidth]{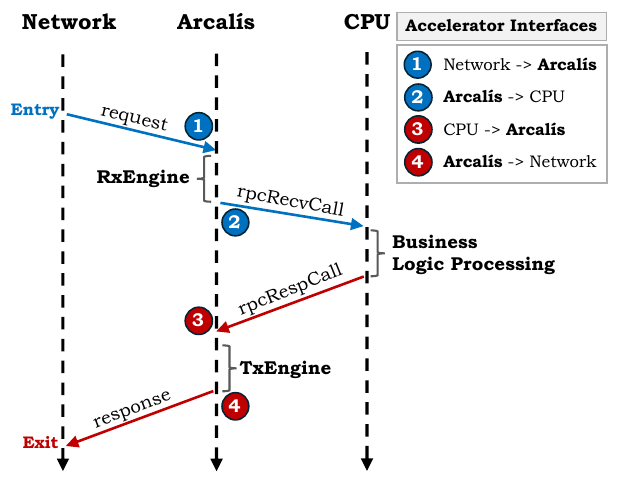}
    \caption{\textit{\Arcalis{} Command Interfaces.}}
    \label{fig:command-interface}
  \end{subfigure}
  \caption{\textit{Architecture of the \Arcalis{} near-cache accelerator and its command interfaces. The fixed logic region (\textbf{FLR}) houses the general-purpose ASICs, while the reconfigurable logic region (\textbf{RLR}) implements the micro-engines.}}
  \label{fig:arcalis-subfigs}
  \vspace{-0.8em}
\end{figure*}

Building on our analysis from Section~\ref{sec:motivation}, we propose \Arcalis{} to eliminate the RPC tax through a full-stack hardware offload while preserving compatibility with existing RPC software frameworks and network transport stacks. \Arcalis{} adopts a philosophy similar to that of the aforementioned recent near-cache accelerators, embedding a cache-coherent compute engine directly within the SoC. Such integration leverages direct cache injection to enable scalable, low-latency RPC acceleration. Our key insight, however, which sets \Arcalis{} apart, is the efficient use of reconfigurable computing to configure an accelerator specifically for the services currently running on the CPU, thereby making a near-cache RPC accelerator space-efficient and practically deployable.


\Arcalis{} is driven by four key architectural goals: (\textit{G1}) completely eliminate CPU involvement in the RPC processing; (\textit{G2}) enable efficient hardware resource utilization; (\textit{G3}) maintain the flexibility to support diverse RPC frameworks and service schemas through software updates alone, without hardware upgrades; and (\textit{G4}) ensure portability and seamless integration with existing hardware mechanisms, RPC runtimes, and networking stacks. 
The remainder of this section details \Arcalis{}' architecture (\S\ref{sec:arcalis-design}), its end-to-end RPC processing pipeline (\S\ref{sec:end-to-end}), and the implementation of \Arcalis{}, including its hardware modeling and software interfaces exposed by microservices to interact with the accelerator (\S\ref{sec:operation_and_design}).

\subsection{Architectural Overview}
\label{sec:arcalis-design}
\Arcalis{} represents a paradigm shift in RPC acceleration through its near-cache hardware accelerator, achieving complete RPC layer offload while maintaining tight integration with the server's memory hierarchy. Figure~\ref{fig:arcalis-architecture} illustrates the comprehensive architecture of \Arcalis{}, which positions itself as a coherent on-chip interconnect component that interfaces directly with both the network infrastructure and CPU cores. Its internal microarchitecture comprises two specialized micro-engines and a shared control block that includes address translation, dispatch, and memory coordination logic. 


We adopt a hybrid architecture comprising a \textit{fixed logic region} (\textit{\textbf{FLR}}) for general-purpose components such as the accelerator's cache \revB{alongside its coherence interface, snooping command interface}, TLB/MMU, load/store queues, address unit, reorder buffer, and command interface, and a \textit{reconfigurable logic region} (\textit{\textbf{RLR}}), containing the offloaded RPC processing logic. 
We select this partitioning to balance efficiency and \revB{adaptability}. FPGAs incur area and performance overheads, so reconfigurability is restricted to the components that benefit most from it. The separation also enhances system stability and security, as the \textit{RLR} operates in the virtual address space while the \textit{FLR} handles physical address management. 
This partition mirrors commercial hybrid architectures: Catapult
separates fixed application-specific integrated circuit (ASIC) logic from reconfigurable fabric~\cite{caulfield2016cloud}, and production eFPGA-augmented SoCs integrate reconfigurable logic
alongside processor cores on the same die~\cite{Lee2022SMIV}.

\Arcalis{} integrates seamlessly into the server's system-on-chip (SoC) design. \revB{The accelerator employs a cache that operates like a core's L2 cache and natively participates in the coherence protocol alongside the other L2 caches and the LLC~\mbox{\cite{arm_amba_chi}}, rather than as an IO-coherent node, which cannot be snooped and therefore cannot cache shareable data~\mbox{\cite{arm_amba_axi}}.} This positioning enables \Arcalis{} to operate as both a network endpoint and a memory-coherent processing unit, \revB{giving cache-line-latency access and direct visibility of CPU writes for} zero-copy data movement between network packets and application buffers, \rev{without the explicit buffer management that DMA-attached accelerators require}. In this paper, we focus on RPC-layer offload and assume transport-layer processing is handled by the network infrastructure.

Next, we discuss the four cardinal design principles around which \Arcalis{} is built. 
Together, they enable efficient, fully offloaded, and application-specialized RPC execution while ensuring seamless integration within modern SoC and network systems. 

\vspace{0.2em}
\noindent\textbf{Complete Hardware Offload:} \hypertarget{complete_offload}{}\Arcalis{} achieves complete hardware offload by implementing all five critical RPC layer modules (\whitecircle{1} - \whitecircle{3}, \whitecircle{5} - \whitecircle{6}) depicted in Figure~\ref{fig:rpc-pipeline}, entirely in hardware, eliminating the need for CPU involvement in the RPC processing pipeline. Unlike software RPC stacks, which incur multiple system calls, kernel networking, and userspace handler overheads~\cite{seemakhupt-sosp23-profiling, kanev-isca15-profilingwsc}, \Arcalis{} eliminates the CPU from the critical path of RPCs by implementing the full control and data flow in hardware.  This approach addresses the fundamental limitation identified in prior works~\cite{pourhabibi-micor21-cerebros, pourhabibi2020optimus}, where partial acceleration creates bottlenecks due to CPU-accelerator coordination overhead. The accelerator maintains four distinct command interfaces as shown in Figure~\ref{fig:command-interface}: two interfaces to the network (interfaces \bluecircle{1} and \winecircle{4}) for request ingress and response egress, and two interfaces to the microservice application (interfaces \bluecircle{2} and \winecircle{3}) for CPU communication. Each interface is equipped with dedicated communication buffers that act as in-cache queues, enabling asynchronous, unblocking data exchange. 

Specifically, \Arcalis{} leverages \textit{uncacheable (\uc{})} 
store operations to deliver task descriptors (e.g., buffer address, payload length, and operation code) to memory-mapped command buffers within the accelerator. A corresponding \uc{} load retrieves completion status or result metadata, enabling low-latency synchronization between the network or application and the accelerator. In our design of \Arcalis{}, we reserve the lowest 4 bits of the physical address, within a designated space, to encode the task opcode (Fig.\ref{fig:cmd_format}), allowing for efficient decoding of the accelerator commands without requiring additional register state or memory transactions. 

\vspace{0.2em}
\noindent\textbf{Pipeline Decoupling:} One of \Arcalis{}’s key design principles is the decoupling of the receive and response pipelines, driven by three observations: first, RPC request and response sizes often differ substantially (e.g., large RPC requests and small RPC responses)~\cite{seemakhupt-sosp23-profiling}; second, responses are often generated asynchronously by the CPU following application execution; and third, overlapping their compute paths within the accelerator enables higher utilization and hides idle cycles by allowing multiple RPCs to be in-flight concurrently, in line with (G2). Accordingly, \Arcalis{} splits processing into two independently schedulable datapaths: the \texttt{RxEngine} and the \texttt{TxEngine}, each optimized for its respective phase of RPC handling. 

\texttt{RxEngine}, activated via interface (\bluecircle{1}), handles ingress processing, including header parsing, function ID dispatch, and payload deserialization. Incoming requests from the NIC are deserialized to determine field boundaries and types. This data is then passed to the appropriate microservice-specific \texttt{recvFunctionN} block. On the other hand, \texttt{TxEngine} is responsible for the response path. When the application (microservice business logic) produces a result, \texttt{TxEngine} is activated via interface (\winecircle{3}).


\vspace{0.2em}
\noindent\textbf{Deployment-Time Reconfigurability:}
Each datacenter server typically runs a limited set of services, often pinned to specific nodes to exploit warm caches~\cite{2021CarreiraFromWarmToHotStart,Luo2021AlibabaMicroservice}.
\Arcalis{} exploits this by implementing only the RPC processing logic required by the active services in the \textbf{\textit{RLR}}, reconfiguring it at deployment time when services change. This reduces area requirements and makes embedding an RPC accelerator in the core die feasible. The
\texttt{recvFunctionN} and \texttt{respFunctionN} blocks shown in Figure~\ref{fig:arcalis-architecture} are the primary targets of reconfiguration: they encode service-specific field
layouts and type expectations derived from the interface description language (IDL) schema, and are replaced when a new service is deployed. 
\textbf{\textit{FLR}} components remain fixed across deployments, as they implement infrastructure independent of any specific RPC schema. 

\vspace{0.2em}
\noindent\textbf{Near-Cache Processing:} \Arcalis{} is connected as a coherent agent on the SoC’s on-chip interconnect, enabling direct and low-latency access to memory and cache hierarchies. \hypertarget{colocation}{}It is physically located near the LLC slices and participates in the memory coherence protocol using LD/ST queues that issue memory operations. These queues are backed by a memory management unit (MMU) and translation lookaside buffer \hypertarget{tlb_consistency}{}(TLB), which is kept consistent with the host TLB and allows \Arcalis{} to translate virtual addresses into physical addresses, enforce access permissions, and support memory access across distinct user-level address spaces, enabling secure and efficient execution. The reorder buffer (ROB) and Address Unit coordinate multiple outstanding memory transactions, ensuring correct sequencing of dependent operations. 

\revPM{We place \Arcalis{} beside the LLC rather than inside the core for three reasons. First, in-core integration is intrusive and does not scale. Prior work has shown that replicating an RPC accelerator at every core of a $64$-core chip costs about $75\%$ of a core's area, versus $\sim$$7\%$ for a single shared instance~\mbox{\cite{pourhabibi-micor21-cerebros}}. 
Second, decoupling \Arcalis{} from the core lets both to be designed and updated independently, so a change to one does not force the redesign of the other. 
Third, inbound network data is commonly injected into the LLC through DCA~\mbox{\cite{intel_ddio, arm_cache_stashing, farshin2020reexamining, yuan2021ioca, kim2025a4}}, so a near-LLC engine is well positioned to consume RPC data. Moreover, this placement operates at cache-line access latencies in the best case while avoiding the overhead of in-core integration.
} 
Compared to PCIe-attached accelerators~\cite{zhang-hpca25-rpcnic}, \Arcalis{} avoids the long PCIe latency overhead. 
It acts as a first-class peer to CPU cores and can access 
data regions with equivalent privileges. This design is especially beneficial for latency-critical microservices, where traditional NIC and kernel stacks introduce tens of microseconds of overhead per call~\cite{sutherland-isca20-nebula, lazarev-asplos21-dagger}. 
\revB{Because \Arcalis{} attaches to the interconnect rather than the LLC pipeline and occupies a separate clock domain, it adds interconnect traffic latency without reducing the LLC frequency.}


\subsection{The End-to-End RPC Execution Pipeline}
\label{sec:end-to-end}
Figure~\ref{fig:end-to-end} illustrates the detailed data and control flow involved in end-to-end RPC execution through \Arcalis{}. The system comprises a network core (\texttt{NetCore}), an application core (\texttt{AppCore}), and the \Arcalis{} accelerator, all connected via a coherent on-chip interconnect and sharing a common LLC. Communication occurs through four shared buffers: \textit{Net. Recv.}, \textit{Net. Resp.}, \textit{App. Recv.}, and \textit{App. Resp.}. 

\begin{figure}[t]
    \centering
    \includegraphics[width=1.0\columnwidth]{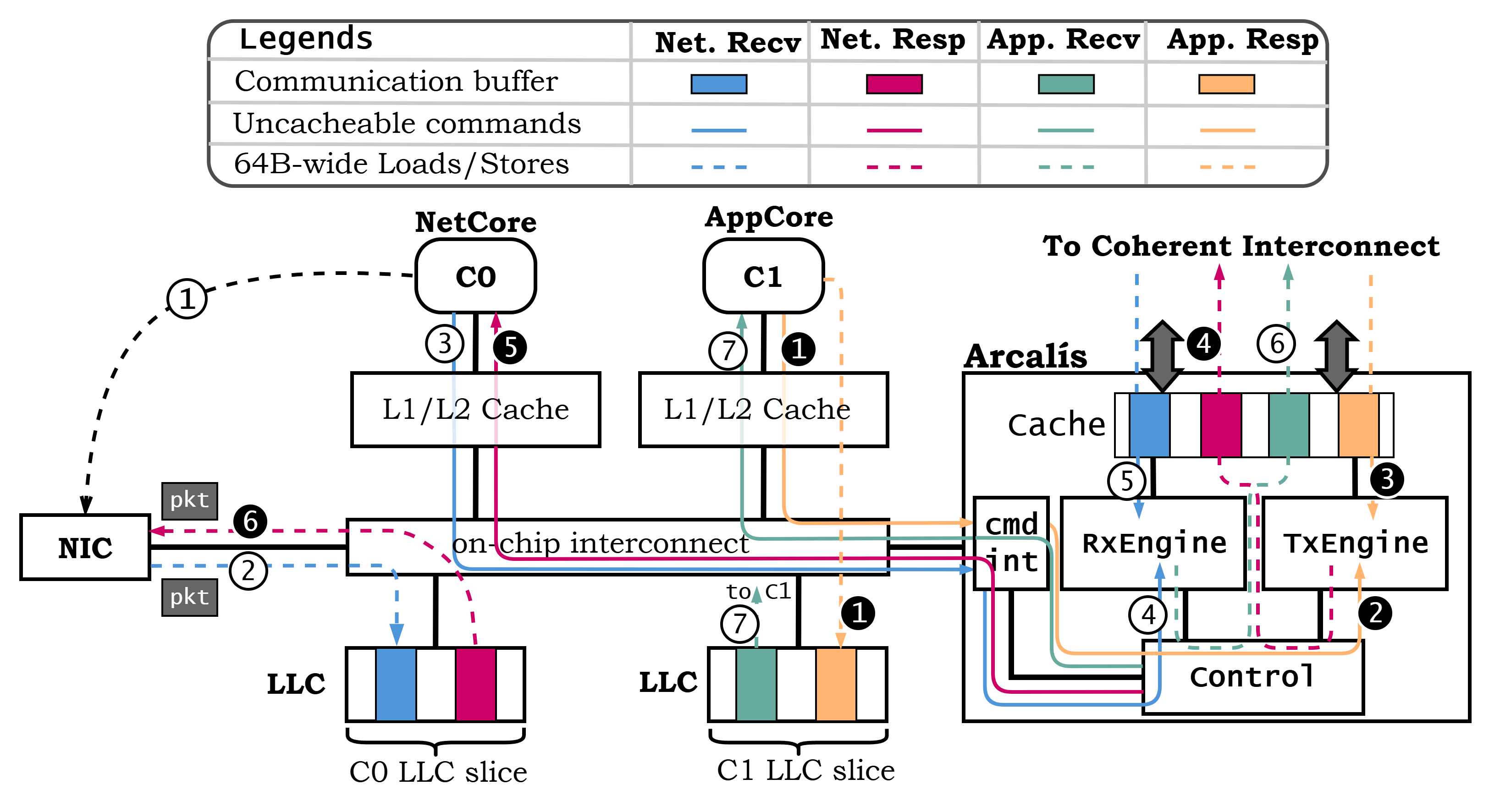}
    \caption{\textit{End-to-end RPC processing through \Arcalis{}.}}
    \label{fig:end-to-end}
    \vspace{-1em}
\end{figure}

If available, the architecture leverages the I/O data placement capabilities of modern CPUs that enable I/O devices to directly place their data into the CPU's caches~\cite{amd_sdci, intel_ddio, arm_cache_stashing}.
Next, we discuss the life cycle of an RPC packet in \Arcalis{}.
\subsubsection{\textbf{Receive Path Processing}}
The receive path (Figure~\ref{fig:end-to-end}) begins at \whitecircle{1} when the \texttt{NetCore} polls the NIC for incoming packets. Upon packet arrival (\whitecircle{2}), the data is DCA-ed from the NIC's on-chip SRAM buffer directly to any of \texttt{NetCore}'s LLC DCA ways, corresponding to the \textit{Net. Recv.} buffer. This DCA-enabled transfer bypasses main memory, ensuring that packet data resides in the cache hierarchy for immediate processing. At \whitecircle{3}, the \texttt{NetCore} issues a \uc{} store to \Arcalis{}, writing a command descriptor containing the address of the packet and relevant metadata. This command is received by the \Arcalis{} control block (\whitecircle{4}), where it is decoded and passed to the central control FSM. The FSM activates the \texttt{RxEngine}, which begins executing the receive-side pipeline.

In \whitecircle{5}, the \texttt{RxEngine} uses the address and length from the command to issue cache-coherent loads to retrieve the packet contents from the \textit{Net. Recv.} buffer. This action is facilitated by the TLB/MMU, and a load queue to issue accesses, while the ROB ensures a correct in-order retirement of memory transactions. The deserializer extracts the RPC function ID, and the dispatch module selects the appropriate \texttt{recvFunctionN} logic block. 

At \whitecircle{6}, the deserialized payload is written into the \textit{App. Recv.} buffer, and the control logic is updated to indicate readiness. In \whitecircle{7}, the \texttt{AppCore} sends a \uc{} load when it is ready to process a request. \Arcalis{} will respond with the \revB{completion} status upon the successful writing of the payload, thereby completing the receive path. The application then begins business logic processing by issuing load requests to retrieve the contents of \textit{App. Recv.} buffer \whitecircle{7}. 

\subsubsection{\textbf{Response Path Processing}}
 Once the microservice function (business logic) completes execution, the \texttt{AppCore} writes 
 its output into the \textit{App. Resp.} buffer (Figure~\ref{fig:end-to-end}). At \darkcircle{1}, it signals \Arcalis{} by issuing a \uc{} store containing the address and metadata of the response payload. The command reaches \Arcalis{} in \darkcircle{2}, where it is decoded and routed to the control logic, which activates the \texttt{TxEngine}. In \darkcircle{3}, the \texttt{TxEngine} accesses the application’s response buffer via cache-coherent loads, using the load queue, ROB, and virtual-to-physical address translation units (TLB/MMU) as in the receive path. Once data retrieval completes, the engine begins header creation and serialization tasks to transform the application data into wire format.

Following RPC processing in \texttt{TxEngine}, the data is written to the \textit{Net. Resp.} buffer in \darkcircle{4}. At \darkcircle{5}, the \texttt{NetCore} issues a \uc{} \revB{load to check whether the response is ready}. \Arcalis{} responds with the completion \revB{status}, as it does to \texttt{AppCore} in the receive path. Finally, in coordination with the networking library and hardware NIC driver, the response is DCA-ed from the \textit{Net. Resp.} buffer to the NIC's on-chip SRAM~\darkcircle{6}, from where the packet response is transmitted over the network, completing the full RPC round-trip.

This end-to-end design enables \Arcalis{} to efficiently process RPCs with zero-copy communication through the cache hierarchy, completely removing the CPU from the latency-sensitive control plane. \Arcalis{} seamlessly integrates with existing RPC load balancing~\cite{seyedroudbari2023turbo, daglis-asplos19-rpcvalet, huang2024hal} and affinity-based steering mechanisms~\cite{pourhabibi-micor21-cerebros}. Unlike \rev{PCIe-attached} accelerators that \rev{incur per-field interconnect traversals during serialization and require CPU-assisted memory pre-staging to mitigate them~\mbox{\cite{zhang-hpca25-rpcnic}}}, \Arcalis{} accesses \rev{application data through the coherent on-chip interconnect at cache-line latency}. This eliminates the need for memory-affinity optimizations and frees \rev{the CPU} entirely for \rev{business} logic execution.

\subsubsection{\textbf{Memory Ordering}}
\revB{Memory ordering between \mbox{\Arcalis{}} and the cores is guaranteed by construction on both paths. The command page carries \uc{} stores that deliver command descriptors and \uc{} loads that return the completion status, while RPC data resides in the separately mapped data pages as subsequently detailed in \S\ref{sec:fixed_logic}.}

\noindent\textbf{\revB{Core-to-\mbox{\Arcalis{}} ordering.}} \revB{A core issues a \uc{} store command whose descriptor points to data in a buffer, and \mbox{\Arcalis{}} must observe that data before acting on the command. On the receive path, after the RPC packet lands in \textit{Net. Recv.}, the \texttt{NetCore} issues the command referencing it (i.e., \whitecircle{3} in Figure~\ref{fig:end-to-end}). On the response path, after business logic execution, the \texttt{AppCore} writes its output into \textit{App. Resp.} and then issues the command referencing it (i.e., \darkcircle{1} in Figure~\ref{fig:end-to-end}). In both cases, correct operation requires only that the referenced data reaches the coherent domain before the \uc{} store command. This holds under total store order (TSO)~\mbox{\cite{x86_tso, riscv_ztso, apple_tso, sparc_tso}}, since stores are not reordered with other stores~\mbox{\cite{intel-sdm-vol3, amd-apm-vol2, Ex86_tso}}. On weakly ordered architectures (i.e., without TSO), a store-release~\mbox{\cite{gharachorloo1990release}} before the command store enforces the required order~\mbox{\cite{arm_arm_ddi0487, riscv_unpriv}}.}

\noindent\textbf{\revB{\mbox{\Arcalis{}}-to-Core ordering.}} \revB{After RPC processing, \mbox{\Arcalis{}} writes results to a buffer and then reports readiness in response to the requesting core's \uc{} load. Before responding, \mbox{\Arcalis{}} enters the DRAIN state and waits until \texttt{MemReqInFlight == 0} (Figure \mbox{~\ref{fig:engine_fsm}}). On the receive path, this ensures the deserialized data written to \textit{App. Recv.} is visible before the \texttt{AppCore} reads the payload for business logic execution. On the response path, it ensures the serialized data in \textit{Net. Resp.} is visible before the \texttt{NetCore} transmits the response.}

\subsection{Operation and Design of \Arcalis{}}
\label{sec:operation_and_design}
\vspace{0.2em}
We now describe the design of our accelerator and how it integrates into other parts of the system. At a high-level, \Arcalis{} comprises the \textit{\textbf{FLR}} and \textit{\textbf{RLR}}, which support its general-purpose operation and microservice-specific needs, respectively. The accelerator integrates into the memory hierarchy and operates alongside the CPU using a state-driven architecture that monitors specific memory regions and a configuration channel. 

\subsubsection{\textbf{Fixed Logic Region}}
\label{sec:fixed_logic}
Forms the ASIC substrate of \Arcalis{}, hosting common components across every microservice, such as the cache, TLB/MMU, load/store queues, address unit, ROB, and command interface. \Arcalis{} operates on two types of memory addresses: (1) The physical address space allocated at initialization for \uc{} communication or ``\textit{\textbf{command page}}'', and (2) The virtual address space used for data exchange or ``\textit{\textbf{data page}}'', allocated by the application. The former is managed in the \textbf{\textit{FLR}}, while the latter is included in the command field (Fig.~\ref{fig:cmd_format}) by the microservice application to initiate requests to be handled in the accelerators' \textbf{\textit{RLR}}. 

\begin{table}[b]
\centering
\vspace{-0.8em}
\caption{\textit{Commands for data exchange with microservices.}}
\vspace{-0.5em}
\label{tab:commands}
\small
\begin{tabularx}{\columnwidth}{|l|X|}
\toprule
\textbf{Command Type} & \textbf{Description} \\
\midrule
\texttt{CMD\_SEND\_NET\_BUF}   & Send network packet buffer address \\
\texttt{CMD\_SEND\_NET\_LEN}   & Send packet length metadata \\
\texttt{CMD\_APP\_READY\_FLAG} & Signal application's readiness for a new RPC data \\
\texttt{CMD\_SEND\_APP\_RESP}  & Send application response data \\
\texttt{CMD\_SEND\_APP\_BUF}   & Send application output buffer \\
\texttt{CMD\_NET\_READY\_FLAG} & Signal network's readiness for a new RPC data \\
\bottomrule
\end{tabularx}
\end{table}

\vspace{0.2em}
\noindent\textbf{Configuration Channel.} 
\Arcalis{} configuration mechanism establishes runtime communication with CPU cores. The OS specifies the physical address range the accelerator monitors for uncacheable (UC) memory requests. Any \uc{} store or load within this range is intercepted and interpreted as an accelerator command. Alternatively, this configuration can be conveyed via PCIe-mapped registers or modern CPU–accelerator primitives such as Intel’s \texttt{ENQCMD}, \texttt{MOVDIR64B}~\cite{yuan-24-intelaia}, or Arm’s System Register Interfaces~\cite{arm_cp14_cp15}. \Arcalis{}' \textbf{\textit{FLR}} design incorporates a \textit{Snooping Command Interface} (\texttt{SCI}) depicted in Figure~\ref{fig:arcalis-architecture} that monitors memory transactions issued by CPU cores. When the kernel driver allocates a \uc{} page for communication, it passes the virtual address to the application and the physical address to \Arcalis{}, with which it programs the traffic snooping watch range. The \texttt{SCI} filters CPU memory operations that are (1) marked \uc{} and (2) fall within the configured range. Moreover, \Arcalis{}' kernel driver pins the \textit{command page}, preventing the OS from changing it. Since configuration occurs only once during initialization, its overhead is negligible.

\Arcalis{} exposes six lightweight control commands shown in Table~\ref{tab:commands} that facilitate communication between user-space microservices and the accelerator through the \uc{} page. We found that these commands were sufficient to capture the communication requirements of representative microservices~\cite{gan-asplos19-deathstarbench}. Each accelerator request is encoded within a 64-bit word, as illustrated in Figure~\ref{fig:cmd_format}: the upper 60 bits carry the data buffer address or length, while the lower 4 bits specify the \textit{OpCode} corresponding to the command type.

\begin{figure}[t]
    \centering
    \includegraphics[width=1\linewidth]{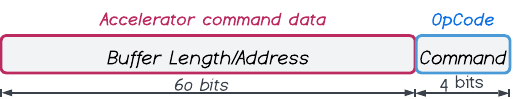}
    \caption{\textit{64-bit accelerator command format in} \Arcalis{}.}
    \label{fig:cmd_format}
   \vspace{-1.2em}
\end{figure}

\vspace{0.2em}
\noindent\textbf{
Memory Management.} \Arcalis{} includes full virtual memory support to reduce access latency and enables direct interaction with data exchange buffers. Address translation operates in 
timing mode to model realistic MMU delays, including TLB misses and page table walks. The page table walker supports 4KB and 2MB data pages. 
A dedicated TLB (Figure~\ref{fig:arcalis-architecture}) accelerates repeated accesses to \rev{the small, fixed set of} hot buffer \rev{pages reused across every RPC request, even though the field data within them exhibit limited reuse.} 
Upon accessing an unmapped address, \Arcalis{}’ translation unit reports a fault to the application. An error code is sent as a response to the pending UC load, so the application retries, triggering the CPU to touch the addresses and the OS to populate the page before \Arcalis{} resumes. The registered buffer regions are pinned allocations, making shootdowns affecting these regions rare in practice. When a shootdown does occur, \rev{the kernel driver, triggered by MMU notifier callbacks~\cite{MMU_notifier}, signals \mbox{\Arcalis{}} via the configuration channel before the OS invalidates the mapping. Then, \mbox{\Arcalis{}} drains all in-flight memory 
operations dependent on the stale mapping before acknowledging the invalidation~\mbox{\cite{2022TLBshootdown}}.} 
\subsubsection{\textbf{Reconfigurable Logic Region}}
\label{sec:reconfigurable_logic}
Hosts the programmable components of \Arcalis{}. Operating in the \textit{virtual address space}, it enables rapid adaptation to evolving RPC frameworks and microservice protocols, as well as facilitates the accelerators' customization for efficiency. 
The \textbf{\textit{RLR}} mainly consists of the \texttt{RxEngine} and \texttt{TxEngine}, as well as their control FSMs. The central FSM in Fig.~\ref{fig:arcalis-architecture} is a simple control block that identifies the command type shown in Table~\ref{tab:commands} and consequently activates the micro-engines for the RPC path execution. Because application needs are dynamic, we position the \texttt{Byte Selector} (selects 4-bit command range) and the response \texttt{Forward Unit} in the \textbf{\textit{RLR}}.

\begin{figure}[t]
\centering
\begin{subfigure}[b]{0.48\columnwidth}
    \centering
    \includegraphics[width=\textwidth]{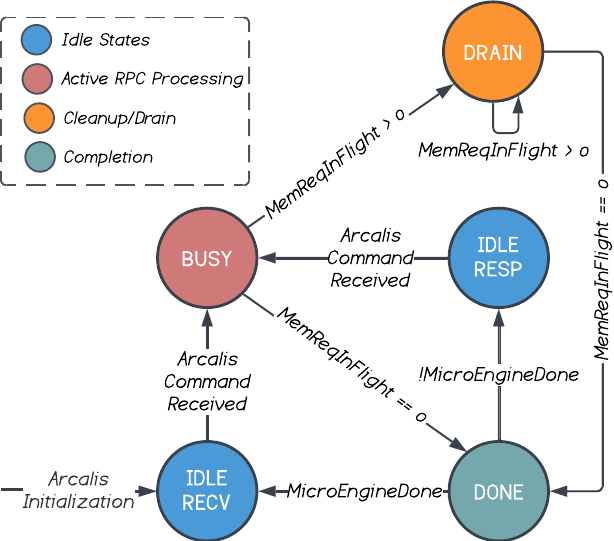}
    \caption{\textit{Micro-Engine FSM}}
    \label{fig:engine_fsm}
\end{subfigure}
\hfill
\begin{subfigure}[b]{0.45\columnwidth}
    \centering
    \includegraphics[width=\textwidth]{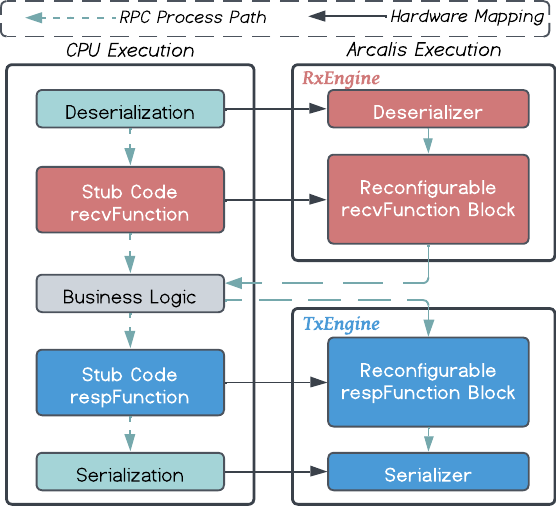}
    \caption{\textit{\Arcalis{} offload model}}
    \label{fig:arcalis-idl-offload}
\end{subfigure}
\caption{\textit{\Arcalis{} accelerator design: (a) state transitions, (b) \Arcalis{} software to hardware mapping.}}
\label{fig:arcalis_implementation}
\vspace{-1em}
\end{figure}

\vspace{0.2em}
\noindent\textbf{State Machine Architecture.}
Figure~\ref{fig:engine_fsm} depicts the \texttt{RxEngine} and \texttt{TxEngine} state machines, which orchestrate deterministic RPC execution across five states. The engine starts in \texttt{IDLE\_RECV} state at initialization. Upon receiving a valid request, it transitions to \texttt{BUSY}, where it performs core RPC operations such as header parsing/creation, serialization via the micro-engines. If outstanding memory operations remain, the engine moves to a separate \texttt{DRAIN} state to ensure all stores complete before finalizing via \texttt{MemReqInFlight}, preventing data loss. Once execution is complete, the engine enters \texttt{DONE} to trigger a response back to the core and cleanup, followed by a transition to either \texttt{IDLE\_RESP} (if the engines still have work to do) or back to \texttt{IDLE\_RECV} to await the next RPC. This fine-grained state separation allows overlapping memory and compute phases, improving pipeline efficiency and RPC throughput. 

\vspace{0.2em}
\noindent\textbf{Specializing IDL-driven (De)Serialization.}
\Arcalis{} offloads both the IDL-generated
stub code and the core RPC data transformation logic to hardware. As shown
in Figure~\ref{fig:arcalis-idl-offload}, the stub code (\texttt{recvFunction}
and \texttt{respFunction}), as well as the (de)serialization operations, are
mapped into the \texttt{RxEngine} on the receive path and the \texttt{TxEngine}
on the response path, executing only the code produced by the IDL compiler.
To make this concrete, consider a Memcached \texttt{SET} request. The IDL
compiler generates a \texttt{recvFunction} that: (1) parses the wire-format
message header to extract field identifiers and lengths, (2) validates each
field against the schema, (3) copies the key and value bytes into a
pre-allocated application buffer, and (4) signals readiness to the
\texttt{AppCore}. In \Arcalis{}, these steps become fixed hardware operations
in the \texttt{RxEngine}: cache-coherent loads from the \textit{Net. Recv.}
buffer, validation against a schema table, and stores to the
\textit{App. Recv.} buffer. Because field types, offsets, and sizes are known
statically from the IDL schema at configuration time, the
\texttt{recvFunctionN} block contains no dynamic dispatch or branch resolution
logic.
Although the current prototype targets Apache Thrift, this IDL-driven approach
is framework-agnostic: any framework that generates stub code from a schema,
including gRPC/protobuf~\cite{grpc} and Cap'n Proto~\cite{capnproto2025}, can
have its stubs mapped to the micro-engines, with reconfiguration occurring once
at deployment time. \Arcalis{} derives its (de)serialization units directly
from the IDL schema, implementing data-movement and format-conversion logic in
the \textbf{\textit{RLR}}, preserving RPC semantics while eliminating the
instruction and control overhead of software-based (de)serialization.

\begin{figure}[t]
  \centering
  \includegraphics[width=\columnwidth]{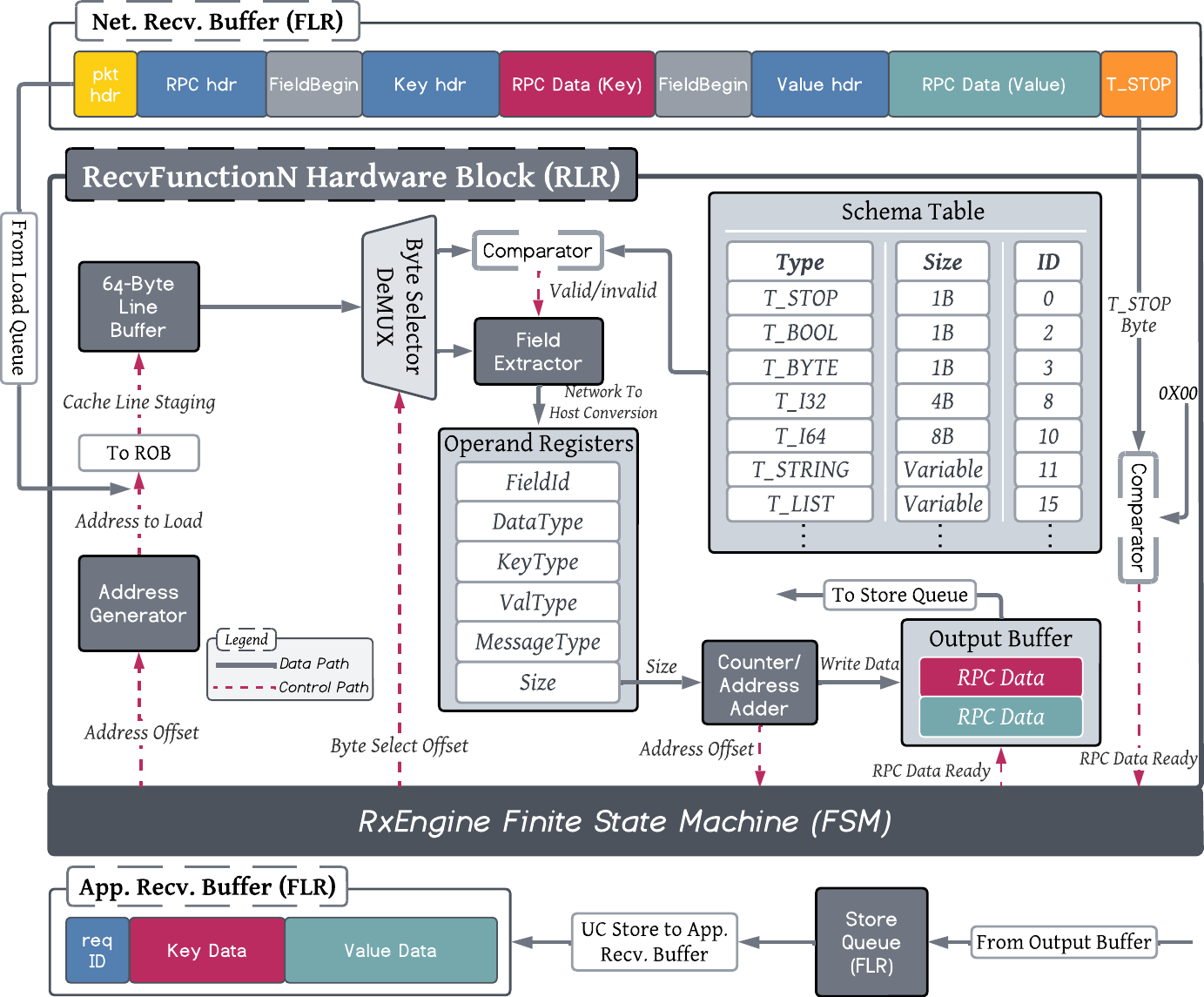}
  \caption{Microarchitecture of \texttt{recvFunctionN} in the \texttt{RLR}, illustrated for $\memcservice$ \texttt{SET} request.}
  \label{fig:recvfunction}
  \vspace{-0.8em}
\end{figure}

The \textbf{Deserializer} in Figure~\ref{fig:arcalis-idl-offload} handles RPC message framing: it parses the message header, extracts the function name, and dispatches to the appropriate \texttt{recvFunctionN} block. Symmetrically, the \textbf{Serializer} writes the RPC message header before \texttt{respFunctionN} assembles the field-level response. Field-level extraction and serialization are handled by the \texttt{recvFunctionN} and \texttt{respFunctionN} blocks respectively, as described below.

\vspace{0.2em}
\noindent\textbf{Designing the Hardware Function Blocks.} Figure~\ref{fig:recvfunction} shows the microarchitecture of the \texttt{recvFunctionN} hardware block, using $\memcservice$ \texttt{SET} request as a
case study.
For each RPC request, \textit{Net. Recv.} buffer holds the payload and its associated metadata (Top of Figure~\ref{fig:recvfunction}). \Arcalis{} inspects the \texttt{pkt hdr} to extract the full payload size (key and value) and then issues all necessary memory requests via the \texttt{FLR} load queue, with response ordering maintained by the ROB.
{Upon arrival of the first cache line, the \textbf{64-Byte Line Buffer} stages it from the ROB.
The \textbf{Byte Selector DeMUX} slices the 512-bit line into header
fields---specifically the field identifier, element type, and
size---which the \textbf{Comparator} validates against the
\textbf{Schema Table}.
The Schema Table is a read-only \texttt{LUT} that encodes the \texttt{RPC}
framework type system: each entry maps a type identifier to its wire
size (e.g., fixed-width scalar or variable-length), and is programmed
once at deployment time for the target framework.
Upon a valid match, the \textbf{Field Extractor} performs
network-to-host byte conversion on the raw header bytes via byte-swap operations.
Validated fields populate the \textbf{Operand Registers}, which hold
per-field metadata including field ID, data type, and size.
The extracted length drives the \textbf{Counter/Address Adder}, which
advances the load address for the next cache line fetch in the ROB.
While header fields are routed through the DeMUX and Comparator,
RPC data (e.g., key and value) bytes are written directly into the Output Buffer as
each cache line arrives.
This loop repeats until the T\_STOP comparator detects the
end-of-message byte~(0x00), at which point the \textbf{RxEngine FSM}
drains the Output Buffer to App. Recv. buffer via the \texttt{FLR} Store Queue (Bottom of Figure~\ref{fig:recvfunction}) and asserts \texttt{ready} to \texttt{AppCore} (step \whitecircle{6} in Figure~\ref{fig:end-to-end}).

The \texttt{respFunctionN} block mirrors this datapath in reverse: it reads structured fields from the App. Resp. buffer, performs host-to-network byte conversion, encodes each field with its type tag and size prefix as specified by the schema, and writes the resulting wire-format bytes to the Net. Resp. buffer via the \texttt{FLR} Store Queue. The hardware footprint of \textit{recvFunctionN} and \rev{\textit{respFunctionN}} blocks scales with \rev{the microservice} schema \rev{complexity (see \S\mbox{\ref{sec:power-area}})}.

\lstset{escapeinside={(*}{*)}}
\begin{lstlisting}[language=C++, caption={\textit{Per-core software integration with Arcalis.}}, label={lst:net_app_core}]
/* NetCore thread */
while (true) {  
  receive_packet(&recv_buf);
  netCallEngineWrite(net_recv_buf, recv_len); 
  netCallEngineRead(net_resp_buf);  
  send_packet(resp_buf);
}
/* AppCore thread */
for (bool done = false; !done;) {  
  appCallEngineRead(app_recv_buf);  
  ComposeUniqueId(); 
  appCallEngineWrite(app_resp_buf, resp_len);
}
\end{lstlisting}
\vspace{-0.8em}
\noindent\textbf{Software Execution Model.} 
Listing~\ref{lst:net_app_core} demonstrates \Arcalis' software execution model where network and application processing are distributed across separate CPU cores, following fundamental specialization principles established in high-performance networking systems~\cite{belay2014ix},~\cite{peter2015arrakis}. The \texttt{NetCore} thread manages packet I/O through DPDK-style interfaces (\textit{receive\_packet}, \textit{send\_packet})~\cite{dpdk} and coordinates with the accelerator via \textit{netCallEngineWrite} and \textit{netCallEngineRead} functions for data buffer exchange. Simultaneously, the \texttt{AppCore} thread focuses exclusively on business logic execution (\textit{ComposeUniqueId}), accessing deserialized RPC data through \textit{appCallEngineRead} and writing responses via \textit{appCallEngineWrite} primitives. The \texttt{NetCore} executes in a continuous loop similar to DPDK's \textit{rte\_eth\_rx\_burst()}~\cite{dpdk}, while the \texttt{AppCore} uses a bounded loop like in Thrift’s \textit{TConnectedClient::run()}\cite{thrift_connected_client}. 
 
While we use \texttt{NetCore} for illustration, \Arcalis{} can operate with hardware-terminated transport~\cite{sutherland-isca20-nebula} or hardware-accelerated NICs~\cite{firestone2018azure} which eliminates the need for a dedicated network core. In such settings, the accelerator can directly interface with the network hardware, further reducing CPU overhead and enabling even tighter integration between network processing and RPC acceleration.

\section{Methodology} 
\begin{table}[t]
\centering
\caption{\textit{Simulation Baseline Configuration.}}
\vspace{-0.5em}
\label{tab:sim-config}
\renewcommand{\arraystretch}{1.2}
\footnotesize
\begin{tabular}{|l|l|}
\toprule
\textbf{Component} & \textbf{Configuration} \\
\midrule
CPU Type & Out-of-Order, 8-wide issue/commit, 4 GHz \\ \hline
Instruction Fetch & 64-entry fetch buffer, 8-wide fetch/decode \\ \hline
Caches & L1I/D: 32 KiB, 8-way, L2: 512 KiB, 8-way \\        
        & L3: 32 MiB (8 slices), 16-way \rev{(MOESI-like)} \\ \hline
Memory & DDR4-2400, 3 GiB, 64-byte interleaving \\ \hline
Topology & 2D Mesh, 1-cycle latency per hop \\ \hline
 \Arcalis{} & \textit{Freq}: 1GHz, \textit{Cache}: 512KiB, 8-way \\
\bottomrule
\end{tabular}
\vspace{-0.8em}
\end{table}

We evaluate \Arcalis{} using the full-system gem5~\cite{binkert-sigarch11-gem5}, extended with a reconfigurable accelerator framework~\cite{Choreographer}, and we implement the accelerator for each evaluated microservice workload. \revB{Our setup faithfully models the full memory hierarchy, where \Arcalis{} participates in coherence as a peer agent on the MOESI-like fabric using its engine cache (\S\ref{sec:arcalis-design}), TLB/MMU translation runs in accurate timing mode, and UC command semantics are modeled, so command-path and address-translation overheads are captured in detail. The separate clock domains in which \Arcalis{} and the CPU cores operate are also modeled, including the interconnect traffic latency.}
The system runs Ubuntu~22.04 with a detailed O3 core for cycle-level execution. The accelerator tile is placed adjacent to the LLC on a 2D mesh NoC. We compare \Arcalis{} against a CPU-only baseline and prior RPC acceleration work, with full parameters shown in Table~\ref{tab:sim-config}. As a proof of concept, we prototyped the \mbox{$\memcservice$} accelerator on an FPGA to validate correctness and report power and area estimates.

\begin{table}[b]
\vspace{-0.8em}
\centering
\caption{\textit{Evaluated microservice workloads. \rev{Low/mid/high denote the write intensity; Values are the operations mix (\%).}}}
\label{tab:workloads}
\setlength{\tabcolsep}{3pt}
\renewcommand{\arraystretch}{1.2}
\scriptsize
\begin{tabularx}{\columnwidth}{|l|X|c|c|c|}
\hline
\textbf{Service} & \textbf{Operations} & \textbf{low} & \textbf{mid} & \textbf{high} \\
\hline
$\memcservice$  & SET / GET                        & 20/80    & 50/50    & 80/20 \\
$\urlservice$   & ComposeUrl / GetExtendedUrl      & 20/80    & 50/50    & 80/20 \\
$\postservice$  & StorePost / ReadPost / ReadPosts & 10/50/40 & 33/33/33 & 90/5/5 \\
$\utservice$    & \{Write/Read\}UserTimeline       & 20/80    & 50/50    & 80/20 \\
$\uidservice$   & ComposeUniqueId                  & \multicolumn{3}{c|}{single operation} \\
\hline
\end{tabularx}
\end{table}

\noindent\textbf{Microservice Workloads \& Configuration.}
We evaluate five microservices from DeathStarBench~\cite{gan-asplos19-deathstarbench} built on Apache Thrift, including $\memcservice$~\cite{memcached}, and $\utservice$ which uses nested RPC. Each offloads RPC parsing and dispatch to \Arcalis{} as described in \S\ref{sec:operation_and_design}. \revB{Following prior work~\cite{pourhabibi-micor21-cerebros}, we study each microservice in isolation to distinguish the RPC acceleration effect from transport and network processing. The NIC is abstracted away in our setup. We replay real-system traces in simulation~\cite{umeike-ispass24-userspace}, placing incoming RPC data directly in host memory as an inbound packet DMA would, so no DCA is modeled and all packet data is fetched from DRAM. The network receive buffer (\textit{Net. Recv.}) sees no cache reuse since each packet arrives at a unique address, while the application buffer (\textit{App. Recv.})  retains some cache warmth from reuse across requests. This is the worst case for cache residency, so our reported results are a lower bound that improves in real deployments where DCA is available.}
We summarize these workloads in Table~\ref{tab:workloads}. To ensure a fair comparison, the CPU baseline processes the same packet traces under exact software and hardware configurations, only with RPC processing executing on the CPU rather than being offloaded to \Arcalis{}. 

\section{Evaluation}

\subsection{End-To-End RPC Acceleration}
\label{sec:end-to-end-evaluation}

Figure~\ref{fig:rpc_bl_breakdown} decomposes execution time into RPC and business logic components. 
\Arcalis{} eliminates nearly all RPC time across all workloads, achieving speedups of 1.72$\times$--4.91$\times$ over the CPU baseline. $\memcservice$ achieves the highest gains, with speedups of 4.91$\times$, 4.05$\times$, and 3.48$\times$ from low to high intensity; speedup decreases with write intensity because \texttt{SET} requests incur higher serialization overhead than \texttt{GET} requests, leaving a larger residual RPC cost. $\postservice$ achieves speedups of 3.41$\times$, 3.14$\times$, and 2.65$\times$, where lower gains reflect greater business logic complexity and a larger non-acceleratable fraction~\cite{gan-asplos19-deathstarbench}. $\urlservice$ yields speedups of 2.51$\times$, 2.25$\times$, and 2.09$\times$, with higher absolute RPC time because of larger message payloads. $\utservice$ shows the widest variance, ranging from 1.72$\times$ to 4.00$\times$; nested RPCs amplify gains at low write intensity, while business logic becomes the bottleneck at high intensity. $\uidservice$ achieves a speedup of 3.88$\times$, confirming \Arcalis{}'s effectiveness for lightweight, RPC-intensive services. 

\begin{figure}[t]
\centering
\includegraphics[width=\columnwidth]{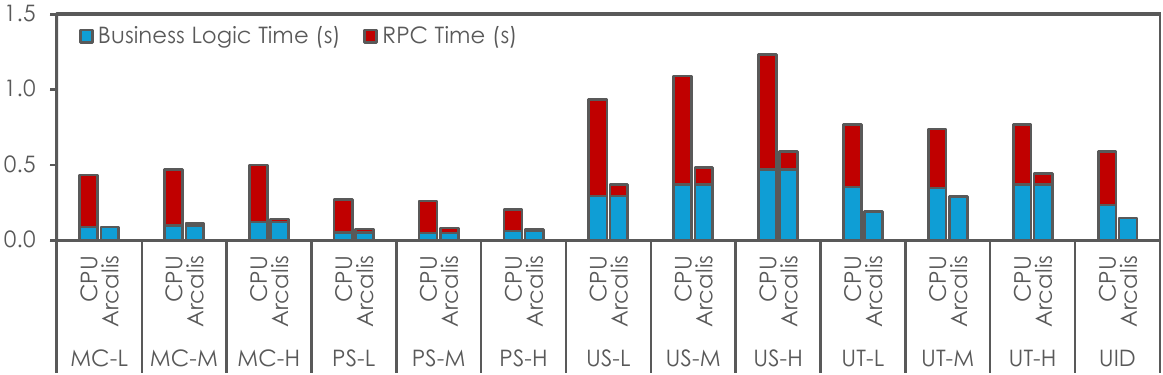}
\caption{\textit{Execution time breakdown into RPC and business logic (BL)
time for CPU and \Arcalis{} across all workloads.
Grouped by service---MC: $\memcservice$,
PS: $\postservice$, US: $\urlservice$,
UT: $\utservice$, UID: $\uidservice$---and by
write intensity: low \rev{(L)}, mid \rev{(M)}, and high \rev{(H)}.}}
\label{fig:rpc_bl_breakdown}
\vspace{-0.8em}
\end{figure}

\begin{figure}[b]
\vspace{-1em}
\centering
\includegraphics[width=\columnwidth]{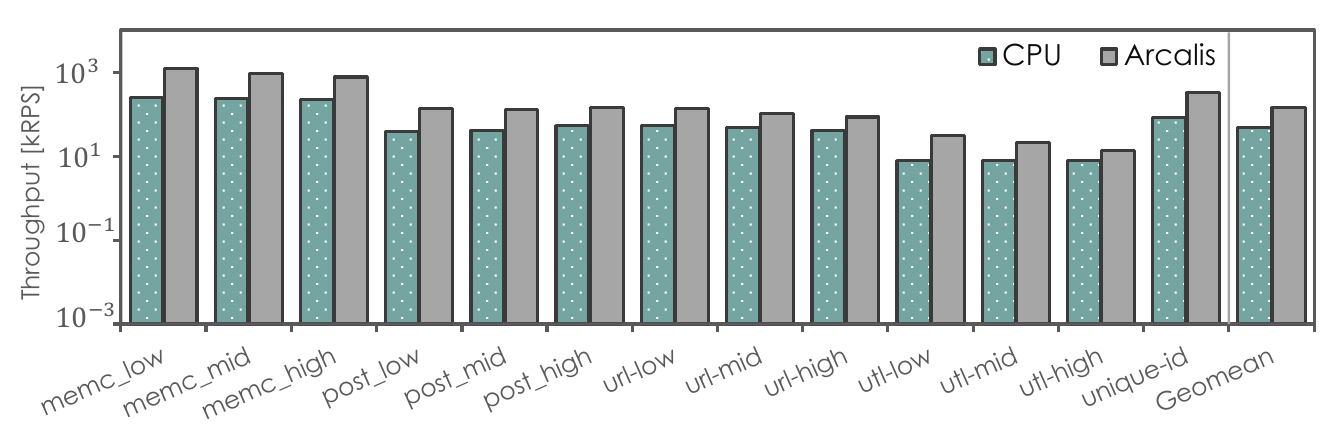}
\caption{\textit{Throughput comparison (kRPS) in log scale between CPU and \Arcalis{} across all workloads.}}
\label{fig:tp_gain}
\end{figure}


Figure~\ref{fig:tp_gain} reports throughput in thousands of requests per second (kRPS). \Arcalis{} achieves 13--1237 kRPS, compared to 8--252 kRPS for the CPU baseline, delivering a geomean throughput improvement of 3.08$\times$. Throughput improvements closely track the reduction in RPC time. $\memcservice$ achieves the highest throughput (762--1237 kRPS), while $\utservice$ yields the lowest absolute throughput of 13--31 kRPS but still results in significant relative gains. This is because its nested RPC structure amplifies the per-hop RPC tax that \Arcalis{} eliminates. These results demonstrate that \Arcalis{} effectively reduces the RPC tax across diverse microservices, with acceleration benefits scaling proportionally to the fraction of execution time spent in RPC processing~\cite{sriraman-iiswc18-microsuite}.

\begin{figure}[t]
   \centering
   \includegraphics[width=1\linewidth]{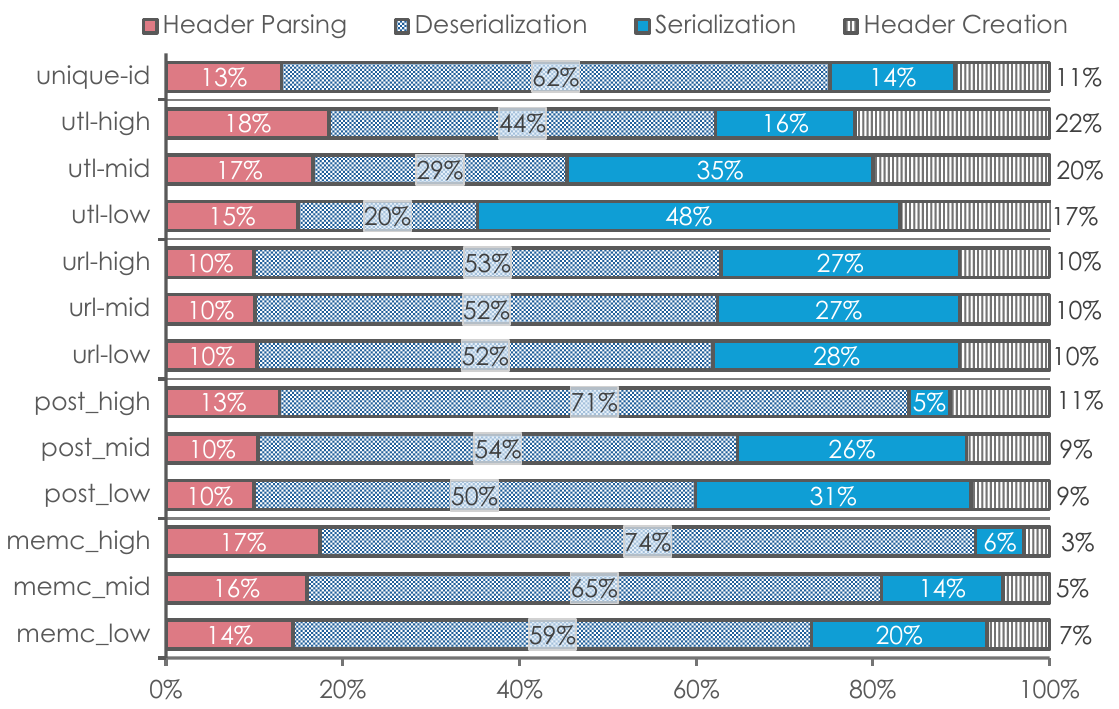}
   \vspace{-1.2em}
   \caption{\textit{Fraction of cycles spent in processing core RPC operations. Dispatch stage is included to Deser. for simplicity.}}
   \label{fig:engine-breakdown}
    \vspace{-0.8em}
\end{figure}

\subsection{\Arcalis{} Processing Time Breakdown}
\label{sec:arcalis_breakdown}
To understand how \Arcalis{} distributes RPC work across its hardware pipeline, Figure~\ref{fig:engine-breakdown} shows the processing time breakdown within the accelerator. \texttt{RxEngine} handles header parsing and deserialization, while \texttt{TxEngine} manages serialization and header creation. $\memcservice$, $\postservice$, and $\uidservice$ are dominated by the deserialization stage at 59\%--74\%, 50\%--71\%, and 62\% respectively. $\urlservice$ shows a stable profile at 52\%--53\% deserialization across all intensities since both of its service functions (\texttt{ComposeUrl} and \texttt{GetExtendedUrls}) involve non-trivial write operations. On the other hand, serialization dominates $\utservice$ for low write intensity workloads at 48\% due to nested RPC response assembly, decreasing to 16\% at high intensity as \texttt{WriteUserTimeline} requests dominate. Across all workloads, write-intensive configurations shift the processing bottleneck toward deserialization for two reasons: (1) fewer read responses reduce serialization work, and (2) write response packets are typically smaller than their requests~\cite{lazarev-asplos21-dagger}. In Figure~\ref{fig:engine-breakdown}, memory operations account for the dominant cost of \Arcalis{}' operation rather than compute (\S\ref{sec:sensitivity}), consistent with the memory-bound nature of RPC microservices (\S\ref{sec:motivation}).

\subsection{Sensitivity Analysis}
\label{sec:sensitivity}

\subsubsection{\textbf{Sensitivity to Interconnect Latency}}
\label{sec:sensitivity-latency}

Figure~\ref{fig:uc-latency} evaluates the impact of accelerator-to-core (command) interconnect latency on end-to-end execution time, using the mid write-intensity configuration for each workload. As latency increases from 5\,ns to 900\,ns, sensitivity correlates inversely with business logic complexity: $\memcservice$ is most sensitive, degrading by up to 114\%, as its short business logic exposes interconnect latency on the critical path. $\urlservice$ and $\uidservice$ follow at 34.6\% and 43.4\%, while $\postservice$ and $\utservice$ are most resilient at 20.9\% and 14.9\%, as their longer business logic masks communication overhead. \revB{This growing degradation as interconnect latency rises shows that RPC workloads, especially latency-critical services like $\memcservice$, suffer when the accelerator sits farther from the core. Since in-core placement is intrusive and does not scale (\S\ref{sec:arcalis-design}), near-cache placement} is critical to realizing acceleration benefits.


\subsubsection{\textbf{Sensitivity to Engine Frequency}}
\label{sec:sensitivity-freq}

Figure~\ref{fig:freq-sensitivity} evaluates how \Arcalis{} engine frequency affects end-to-end performance across a range of 250\,MHz to 1333\,MHz, measured as the percentage improvement over the baseline 1\,GHz configuration. Performance variation across all workloads remains within $\pm$3.2\% across the entire frequency range, with most workloads showing less than 1\% sensitivity. This insensitivity arises because RPC processing is memory-bound rather than compute-bound (\S\ref{sec:motivation}), so engine compute is not on the critical path. $\uidservice$ is the most sensitive at $-$3.2\% at 250\,MHz due to its single-operation business logic (\texttt{ComposeUniqueID}), but recovers to $+$0.6\% at 1333\,MHz. Though not shown, this trend is true even at higher frequencies of 2000-4000\,MHz. These results demonstrate that \Arcalis{} is not sensitive to engine frequency, making it practically deployable as a hybrid architecture (\S\ref{sec:arcalis-design}).

\begin{figure}[t]
\centering
\includegraphics[width=\columnwidth]{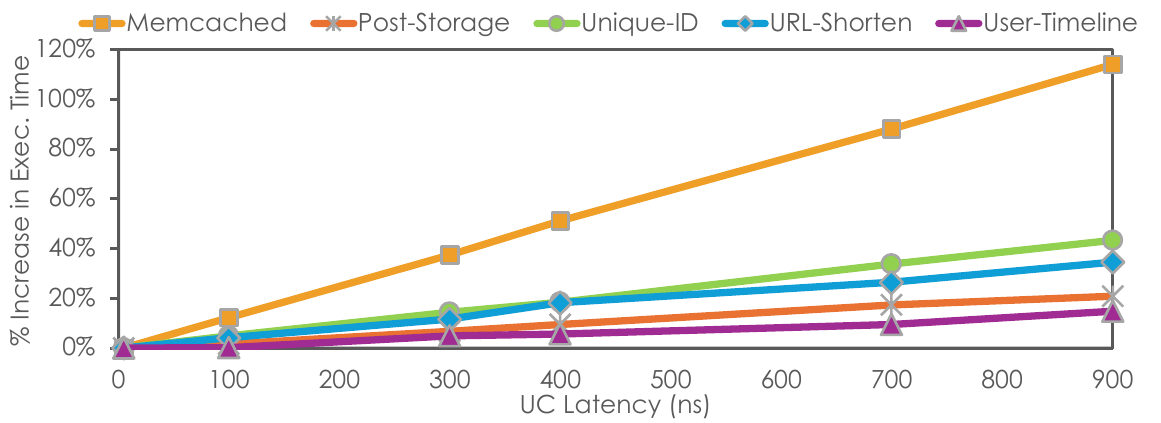}
\vspace{-1.2em}
\caption{\textit{Sensitivity of end-to-end execution time to accelerator-to-core interconnect latency.}}
\label{fig:uc-latency}
\vspace{-0.8em}
\end{figure}

\begin{figure}[b]
\vspace{-1.5em}
\centering
\includegraphics[width=\columnwidth]{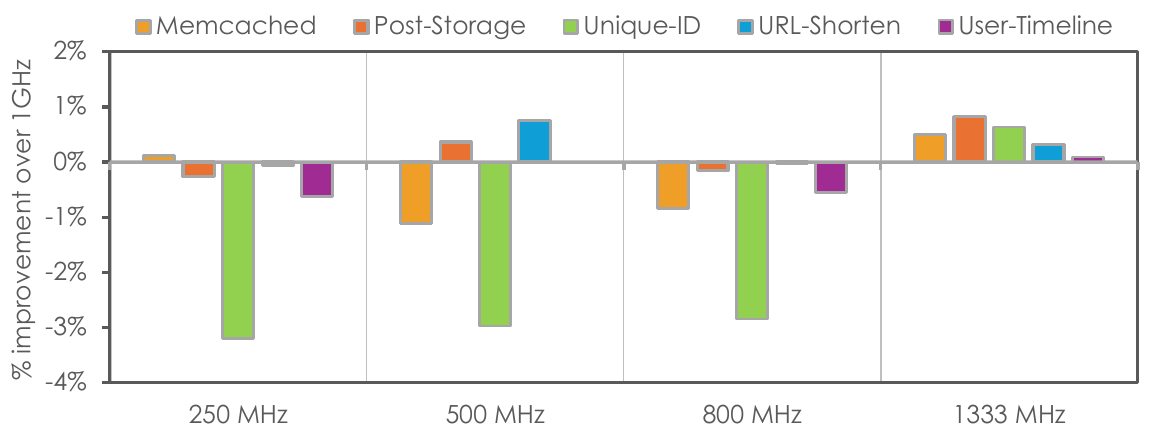}
\vspace{-1em}
\caption{\textit{Sensitivity of end-to-end execution time to \Arcalis{} engine frequency over 1\,GHz baseline.}}
\label{fig:freq-sensitivity}
\end{figure}

\begin{figure*}[t]
    \centering
    \vspace{-1em}
    \includegraphics[width=1\linewidth]{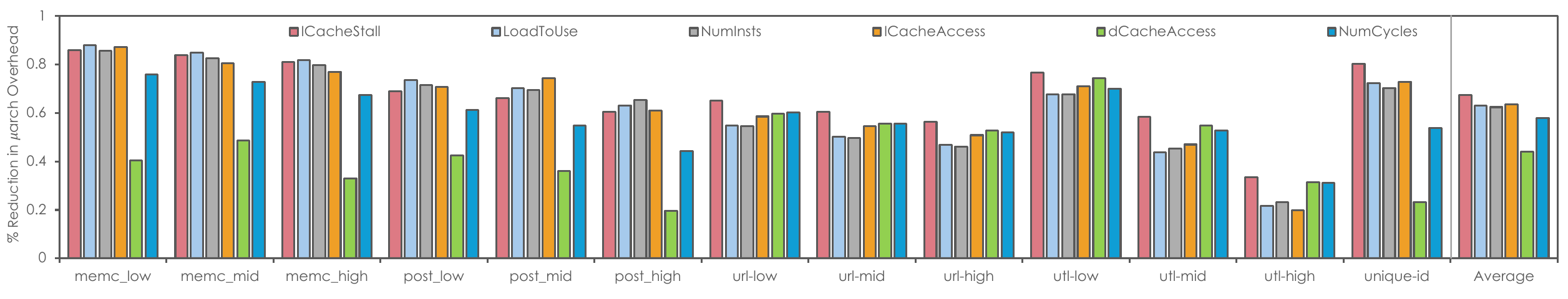}
    \vspace{-1.8em}
    \caption{\textit{CPU microarchitectural overhead reduction in instruction count, cache stalls, cache accesses, and cycles with \Arcalis{}.}}
    \label{fig:microarch-reduction}
    \vspace{-1em}
\end{figure*}

\subsection{Microarchitectural Analysis}
\label{sec:microarch-impact}

Figure~\ref{fig:microarch-reduction} quantifies the reduction in CPU pipeline slot overhead achieved by \Arcalis{} by comparing instruction count, cycle count, cache accesses, and stall rates between the CPU baseline and \Arcalis{}, complementing the characterization in Figure~\ref{fig:vtune_profile} where up to 52\% of slots are lost to front-end and back-end stalls in RPC-intensive workloads. 
On average, instruction count drops by 62.4\% and cycle count by 57.8\%, confirming that offloaded RPC work directly reduces CPU execution burden. Instruction cache accesses fall by 63.5\% and stalls by 67.5\%, driven by the large instruction working sets of RPC processing in microservices~\cite{pourhabibi-micor21-cerebros}, while LoadToUse latency improves by 63.1\%, indicating fewer dependencies and a shorter critical path once RPC handling is offloaded. Data cache accesses decrease by 44.0\% on average; write-intensive workloads generally see smaller reductions compared to read-intensive workloads as business logic operations dominate residual data cache activity, with \textit{post\_high} and \textit{utl\_high} reaching as low as 19.6\% and 31.4\%. $\utservice$ sees the largest data cache access reduction at low write intensity (74.5\%), as nested RPC calls amplify memory access offload, dropping to 31.4\% at high intensity. Overall, \Arcalis{} substantially reduces front-end and back-end overheads across all workloads, freeing CPU resources for business logic.

\begin{figure}[b]
   \centering
   \vspace{-1.2em}
   \includegraphics[width=1\linewidth]{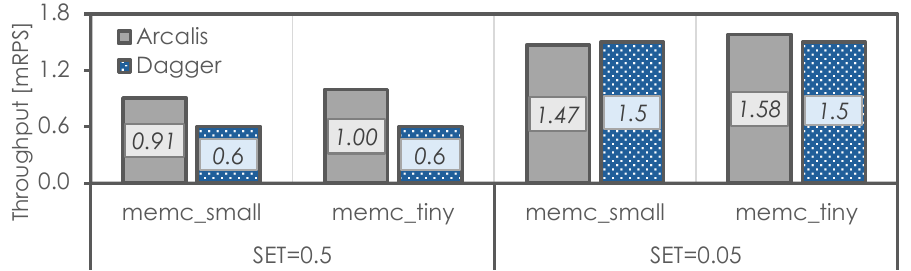}
   \caption{\textit{Throughput comparison between \Arcalis{} and Dagger on $\memcservice$ microservice.}} 
   \label{fig:arcalis-vs-dagger}
   \vspace{-1em}
\end{figure}

\vspace{-0.1em}
\subsection{Comparison to Prior RPC Accelerators}
\label{sec:comparison}
Among the \revB{reconfigurable} prior works in Table~\ref{tab:rpc_table}, Dagger and RpcNIC are the most relevant comparisons to \Arcalis{}. Figure~\ref{fig:arcalis-vs-dagger} compares \Arcalis{} with Dagger~\cite{lazarev-asplos21-dagger} under $\memcservice$ workloads using the same key–-value sizes and SET/GET mixes. For SET=0.5, \Arcalis{} achieves 1.52$\times$ and 1.67$\times$ higher throughput on \textit{memc\_small} and \textit{memc\_tiny}, respectively. At SET=0.05, the two systems perform similarly at 0.98$\times$--1.05$\times$ throughput, as GET-dominated request traffic with small payloads requires minimal
deserialization work, shifting the bottleneck to the memcached lookup that neither system offloads. At SET=0.5, write  traffic dominates, and \Arcalis{}' near-LLC placement eliminates the data movement overhead that Dagger
incurs over its UPI link. \Arcalis{} thus provides the greatest gains precisely where production RPC workloads spend most of their time---write-dominant methods~\cite{seemakhupt-sosp23-profiling}.

RpcNIC~\cite{zhang-hpca25-rpcnic} offers another point of comparison. It reports a speedup of $\sim$1.4$\times$ on both $\uidservice$ and $\urlservice$, while \Arcalis{} reaches 3.88$\times$ and 2.51$\times$, respectively. The higher gains result from performing RPC parsing and dispatch in the near-cache region and avoiding PCIe traversal, which lowers offload overheads without the need for any specialized memcpy engines~\cite{intel_dsa, kuper2024quantitative}. 

\subsection{Power and Area Analysis}
\label{sec:Power and Area}


\vspace{-0.3em}

\label{sec:power-area}

\rev{We estimate \mbox{\Arcalis{}} area and dynamic power across the \texttt{RLR}
and \texttt{FLR} for three services spanning simple to complex RPC schemas,
and quantify the resulting system overhead.}

\begin{figure}[t]
    \centering
    \begin{subfigure}[t]{0.515\linewidth}
        \centering
        \includegraphics[width=\linewidth]{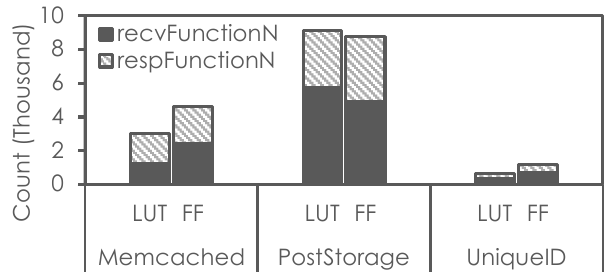}
        \caption{\rev{Resource utilization.}}
        \label{fig:rlr-util}
    \end{subfigure}
    \begin{subfigure}[t]{0.47\linewidth}
        \centering
        \includegraphics[width=\linewidth]{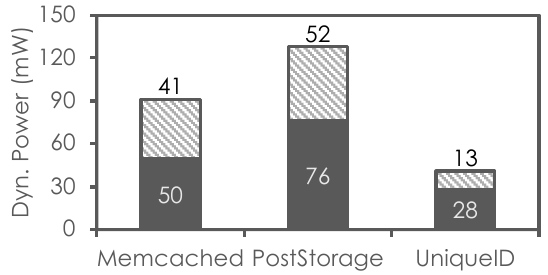}
        \caption{\rev{Dynamic power (mW).}}
        \label{fig:rlr-power}
    \end{subfigure}
    \caption{\rev{\textit{RLR resource utilization and dynamic power, showing recvFunctionN and respFunctionN breakdown.}}}
   \vspace{-1.2em}
    \label{fig:rlr-area-power}
\end{figure}

\subsubsection{\textbf{Reconfigurable Logic Region}}
\rev{Figure\mbox{~\ref{fig:rlr-util}} and Figure\mbox{~\ref{fig:rlr-power}} report \texttt{RLR} resource utilization and dynamic power per service, respectively, after Vivado place-and-route at 280~MHz on the \texttt{xa7a12t-cpg238-2I}. $\postservice$ incurs the highest hardware complexity at 9,115~LUTs, 8,760~FFs, and 128~mW as schema field count and nesting depth in the IDL definition jointly determine \texttt{RLR}'s footprint rather than the service call depth. $\uidservice$ is the most compact at 631~LUTs, 1,182~FFs, and 41~mW. The receive path dominates \texttt{RLR} logic across services, reflecting the \textit{RxEngine} cycle dominance observed in \S\mbox{\ref{sec:arcalis_breakdown}}}.

\rev{As described in \S\mbox{~\ref{sec:arcalis-design}}, the \texttt{RLR}
targets an embedded FPGA fabric. To estimate the silicon area of our accelerator designs, we derive
an LUT density from the AMD VU19P FPGA, which contains 4,086K~LUTs on a
900~mm\textsuperscript{2} die~\mbox{\cite{vu19p,vu19p_die_size}}, yielding $\sim$5,000~LUT/mm\textsuperscript{2}
after accounting for unused DSPs, BRAMs and IOs. At this density,
$\postservice$ occupies an area of $\sim$1.82~mm\textsuperscript{2}. Optimized embedded
FPGA fabric\mbox{~\cite{Adicsys,flex-logix}} reduces the routing overhead and achieves up to 10,000~LUT/mm\textsuperscript{2} at 16~nm,
reducing $\postservice$ to 0.91~mm\textsuperscript{2} and $\uidservice$ to
0.06~mm\textsuperscript{2}. Even at the conservative estimate, the worst-case
\texttt{RLR} sizing remains within the footprint of a single CPU core,
while using 19$\times$--31$\times$ fewer logic resources compared to State-of-the-Art FPGA-based RPC accelerators\mbox{~\cite{zhang-hpca25-rpcnic, lazarev-asplos21-dagger}}.}

\vspace{0.2em}
\subsubsection{\textbf{Fixed Logic Region}}
\rev{Table\mbox{~\ref{tab:flr}} reports \texttt{FLR} area and dynamic power at 16~nm, estimated using McPAT\mbox{~\cite{mcpat}} for memory structures, CACTI\mbox{~\cite{cacti}} for the engine cache, and Vivado synthesis scaled to 16~nm for the control FSM~\mbox{\cite{kuon2006}}. Engine cache dominates at 78\% of area and 76\% of dynamic power, consistent with prior observations that SRAM structures dominate area in data-path accelerators~\mbox{\cite{Lee2022SMIV}}. \revB{The FLR area and power include the coherence interfaces.} The total \texttt{FLR} footprint is 0.464~mm\textsuperscript{2} with 46~mW dynamic power.}

\begin{table}[t]
\centering
\small
\caption{\rev{\textit{FLR area and dynamic power composition.}}}
\vspace{-0.4em}
\label{tab:flr}
\begin{tabular}{l|r|r|r|r|r}
\hline
\textbf{Component} & Cache & LSQ & FSM & DTLB & Others \\
\hline
\textbf{Area}          & 78\% & 12\% & 4\% & 3\% & 3\% \\
\textbf{Dynamic Power} & 76\% & 18\% & 3\% & 2\% & 1\% \\
\hline
\end{tabular}
\vspace{-1.2em}
\end{table}


\rev{An 8-core baseline system occupies approximately
50.93~mm\textsuperscript{2} at 16~nm, derived from the 2-core SMIV
area of 5.88~mm\textsuperscript{2}~\mbox{\cite{Lee2022SMIV}} and
an SRAM density of 8.1~Mb/mm\textsuperscript{2}~\mbox{\cite{ultra_dense_sram}},
with a single \mbox{Cortex-A53} core occupying
2.42~mm\textsuperscript{2} including private caches.
Combining \textit{FLR} and worst-case \textit{RLR} at conservative
FPGA density, \mbox{\Arcalis{}} requires 2.28~mm\textsuperscript{2}---
approximately one CPU core. At the higher eFPGA density achievable
with optimized embedded fabric~\mbox{\cite{Adicsys}}, \mbox{\Arcalis{}}
shrinks to 1.37~mm\textsuperscript{2}, or 0.57$\times$ of a single
core.}

\vspace{0.2em}
\subsubsection{\textbf{Energy Efficiency}}
\rev{To quantify \mbox{\Arcalis{}'} energy efficiency, we use \mbox{MCPAT~\cite{mcpat}}
to estimate CPU dynamic power from the gem5 simulation in
Table\mbox{~\ref{tab:sim-config}}, yielding 1.72~W for the CPU-only
baseline and 1.20~W when running alongside \mbox{\Arcalis{}} using a 16~nm process.
By offloading the full RPC processing, including header parsing, header creation, (de)serialization, and
dispatch to a dedicated near-cache logic while retaining the business
logic on the CPU, \mbox{\Arcalis{}} reduces the per-request energy from
7.4~$\mu$J, 39.1~$\mu$J, and 20.5~$\mu$J on the CPU-only baseline to 1.3~$\mu$J, 7.5~$\mu$J,
and 4.2~$\mu$J for $\memcservice$, $\postservice$, and $\uidservice$,
respectively. This accounts for both the 174~mW \mbox{\Arcalis{}} tile and the reduced CPU dynamic power of
1.20~W, which is a 5.8$\times$, 5.2$\times$, and 4.9$\times$ system energy reduction.}

\section{Related Work}
\noindent\textbf{Hybrid ASIC and Reconfigurable Architectures.}
Some prior works have explored the feasibility of combining fixed-function ASICs with reconfigurable logic. Microsoft's Catapult~\cite{caulfield2016cloud} establishes the design philosophy of separating stable infrastructure (PCIe, memory controllers, network transport) from reconfigurable fabric that handles application-specific processing. SMIV~\cite{Lee2022SMIV} demonstrates a 16nm SoC integrating ARM cores with an eFPGA and coherent accelerators on the same die, with the eFPGA achieving 1~GHz operation. Commercial products, such as Zynq~\cite{Zynq7000}, eFPGA IP from Flex Logix~\cite{flexlogix}, and QuickLogic~\cite{quicklogic2021efpga} confirm that embedding reconfigurable fabric within ASIC designs is viable at production scale. \Arcalis{} adopts this same principle: general-purpose components reside in the \texttt{FLR} as ASIC, while service-specific RPC datapaths occupy the \texttt{RLR} and are reconfigured \rev{only} at deployment time \rev{for a specific microservice}. 
\vspace{0.2em}

\noindent\textbf{Reconfigurability and RPC Acceleration.}
\revB{Prior on-chip RPC accelerators are schema-programmable but fix their datapath in silicon\mbox{~\cite{pourhabibi-micor21-cerebros, pourhabibi2020optimus, karandikar-micro21-protoacc}}, so keeping pace with evolving RPC frameworks eventually requires a hardware change, motivating the reconfigurable approach \mbox{\Arcalis{}} adopts. Partial reconfiguration via the internal configuration access port (ICAP) achieves 6.4\,Gb/s bandwidth\mbox{~\cite{icap}}, enabling RLR reconfiguration in under 1\,ms, well below service deployment latency and entirely off the per-request critical path. Unlike prior RPC accelerators that place their offload at PCIe-attached NICs\mbox{~\cite{zhang-hpca25-rpcnic}}, on-chip NICs at the chip's edge\mbox{~\cite{pourhabibi-micor21-cerebros, sutherland-isca20-nebula}}, or across a separate NUMA interconnect\mbox{~\cite{lazarev-asplos21-dagger}}, \mbox{\Arcalis{}} is the first to combine near-LLC coherent placement, full RPC-layer offload, and deployment-time service-specific reconfigurability. ARM cache stashing\mbox{~\cite{arm_cache_stashing}}, Intel DDIO\mbox{~\cite{intel_ddio}}, and AMD SDCI\mbox{~\cite{amd_sdci}} enable NICs to inject packet data directly into reserved LLC ways, providing LLC residency for incoming packets. \mbox{\Arcalis{}} is also naturally positioned to benefit from such datacenter I/O optimization mechanisms.}

\vspace{0.2em}
\noindent\textbf{Near-Memory and Cache-Adjacent Computing.}
Near-memory computing has gained traction for data-intensive workloads~\cite{kim2017toward, singh2018review}. Processing-in-memory (PIM) designs like Samsung's HBM-PIM~\cite{lee2021hardware} and UPMEM~\cite{mutlu2022modern} place compute units within memory modules, while near-data accelerators ~\cite{denzler-ieee23-casper, nori-isca21-reduct, wang-hpca22-nearstream, izmirli-cal25-halis} position a specialized logic adjacent to the cache. Cache-oriented designs include directory caches~\cite{valls2015ps}, near-cache prefetchers~\cite{heirman2018near, deb2019performance, nguyen2026pickle}, and coherence accelerators~\cite{boroumand2019conda}. However, these approaches target compute-intensive kernels or memory bandwidth optimization rather than protocol processing. \Arcalis{} extends this paradigm by demonstrating that near-cache placement effectively addresses RPC's memory access latency and cache inefficiencies.

\section{Conclusions}
\label{sec:conclusion}

This paper presents \Arcalis{}, a near-cache RPC accelerator that addresses the RPC tax in modern microservices by positioning lightweight, reconfigurable hardware engines adjacent to the last-level cache. By fully offloading RPC parsing, deserialization, serialization, and dispatch to dedicated microengines that operate at cache-line latency, \Arcalis{} eliminates CPU involvement in the RPC critical path while preserving programmability through deployment-time reconfiguration. Evaluated across five representative microservice workloads spanning diverse operation mixes and RPC complexities, \Arcalis{} achieves 1.72--4.91$\times$ end-to-end speedups, reduces microarchitectural overhead by up to 88\%, delivers up to 1.62$\times$ higher throughput than prior solutions, and \rev{cuts per-request energy by 4.9$\times$--5.8$\times$ compared to the CPU-only baseline, all within a practical 1.37--2.28\,mm\textsuperscript{2} silicon footprint}.
These results demonstrate that near-cache placement, combined with service-specific reconfigurability, is a practical and deployable path to overcoming the RPC tax in high-performance microservice deployments.

\section*{Acknowledgments}
\vspace{-0.1em}
\textit{Johnson Umeike is supported by Bahar Asgari’s NSF PPoSS program award under Award Number 2316177. This paper is also part of a project supported by the U.S. Department of Energy (DOE), Office of Science, Advanced Scientific Computing Research (ASCR), under the Early Career Research Program award DE-SC0024079, awarded to Bahar Asgari.}

\bibliographystyle{IEEEtran}
\bibliography{refs}

@inproceedings{kanev-isca15-profilingwsc,
author = {Kanev, Svilen and Darago, Juan Pablo and Hazelwood, Kim and Ranganathan, Parthasarathy and Moseley, Tipp and Wei, Gu-Yeon and Brooks, David},
title = {Profiling a warehouse-scale computer},
year = {2015},
isbn = {9781450334020},
publisher = {Association for Computing Machinery},
address = {New York, NY, USA},
url = {https://doi.org/10.1145/2749469.2750392},
doi = {10.1145/2749469.2750392},
booktitle = {Proceedings of the 42nd Annual International Symposium on Computer Architecture},
pages = {158–169},
numpages = {12},
location = {Portland, Oregon},
series = {ISCA '15}
}

@inproceedings{seemakhupt-sosp23-profiling,
author = {Seemakhupt, Korakit and Stephens, Brent E. and Khan, Samira and Liu, Sihang and Wassel, Hassan and Yeganeh, Soheil Hassas and Snoeren, Alex C. and Krishnamurthy, Arvind and Culler, David E. and Levy, Henry M.},
title = {A Cloud-Scale Characterization of Remote Procedure Calls},
year = {2023},
isbn = {9798400702297},
publisher = {Association for Computing Machinery},
address = {New York, NY, USA},
url = {https://doi.org/10.1145/3600006.3613156},
doi = {10.1145/3600006.3613156},
booktitle = {Proceedings of the 29th Symposium on Operating Systems Principles},
pages = {498–514},
numpages = {17},
location = {Koblenz, Germany},
series = {SOSP '23}
}

@misc{MMU_notifier,
    title = {Heterogeneous memory management and {MMU} notifiers},
    author={Jonathan Corbet},
    howpublished = {\url{https://www.amd.com/content/dam/amd/en/documents/epyc-technical-docs/white-papers/58725.pdf}},
   note = {Accessed: 2025-7-25}
}

@misc{amd_sdci,
  title = {{S}mart {D}ata {C}ache {I}njection ({SDCI}) White Paper
},
  author={AMD},
  howpublished = {\url{https://www.amd.com/content/dam/amd/en/documents/epyc-technical-docs/white-papers/58725.pdf}},
  note = {Accessed: 2025-7-25}
}

@misc{flexlogix,
  title = {Flex Logix High-Performance Embedded FPGA IP Core 
},
  author={EDN},
  year = {2016},
  howpublished = {\url{https://www.edn.com/flex-logix-high-performance-embedded-fpga-ip-core-now-available-for-tsmc-16ff-and-16ffc/}},
  note = {Accessed: 2025-11-15}
}

@misc{arm_cache_stashing,
 author = {{Arm Limited}},
 title = {{Cache stashing}},
 year = {2025},
 url = {https://developer.arm.com/documentation/102407/0102/Cache-stashing},
 note = {Arm Developer Documentation, Accessed: 2025-07-26}
}

@techreport{intel_ddio,
 author = {{Intel Corporation}},
 title = {{Intel Data Direct I/O Technology}},
 institution = {Intel Corporation},
 year = {2025},
 type = {Technology Brief},
 url = {https://www.intel.com/content/dam/www/public/us/en/documents/technology-briefs/data-direct-i-o-technology-brief.pdf},
 note = {Accessed: 2025-07-26}
}

@misc{Choreographer,
      title={Choreographer: A Full-System Framework for Fine-Grained Tasks in Cache Hierarchies}, 
      author={Hoa Nguyen and Pongstorn Maidee and Jason Lowe-Power and Alireza Kaviani},
      year={2025},
      eprint={},
      archivePrefix={arXiv},
      primaryClass={},
      url={https://arxiv.org/abs/2510.26944}, 
}

@inproceedings{Duplyakin+:ATC19,
    title     = "The Design and Operation of {CloudLab}",
    author    = "Dmitry Duplyakin and Robert Ricci and Aleksander Maricq and Gary Wong and Jonathon Duerig and Eric Eide and Leigh Stoller and Mike Hibler and David Johnson and Kirk Webb and Aditya Akella and Kuangching Wang and Glenn Ricart and Larry Landweber and Chip Elliott and Michael Zink and Emmanuel Cecchet and Snigdhaswin Kar and Prabodh Mishra",
    booktitle = "Proceedings of the {USENIX} Annual Technical Conference (ATC)",
    pages     = "1--14",
    year      = 2019,
    month     = jul,
    url       = "https://www.flux.utah.edu/paper/duplyakin-atc19"
}

@misc{dpdk,
 author = {{Data Plane Development Kit}},
 title = {{DPDK.org}},
 year = {2025},
 url = {https://www.dpdk.org/},
 note = {Accessed: 2025-07-26}
}

@misc{thrift,
 author = {{Apache Thrift Software Framework}},
 title = {{thrift.org}},
 year = {2025},
 url = {https://thrift.apache.org/},
 note = {Accessed: 2025-07-26}
}

@misc{memcached,
 author = {{Memcached}},
 title = {{Memcached - a distributed memory object caching system}},
 year = {2025},
 url = {https://github.com/memcached/memcached},
 note = {Accessed: 2025-07-26}
}

@misc{thrift_connected_client,
 author = {{Apache Software Foundation}},
 title = {{TConnectedClient.cpp}},
 year = {2025},
 url = {https://github.com/apache/thrift/blob/master/lib/cpp/src/thrift/server/TConnectedClient.cpp#L55},
 note = {Line 55, Accessed: 2025-07-26}
}

@misc{grpc,
 author = {{gRPC}},
 title = {{gRPC - A high performance, open source universal RPC framework}},
 year = {2025},
 url = {https://grpc.io/},
 note = {Accessed: 2025-07-26}
}

@inproceedings{gan-asplos19-deathstarbench,
author = {Gan, Yu and Zhang, Yanqi and Cheng, Dailun and Shetty, Ankitha and Rathi, Priyal and Katarki, Nayan and Bruno, Ariana and Hu, Justin and Ritchken, Brian and Jackson, Brendon and Hu, Kelvin and Pancholi, Meghna and He, Yuan and Clancy, Brett and Colen, Chris and Wen, Fukang and Leung, Catherine and Wang, Siyuan and Zaruvinsky, Leon and Espinosa, Mateo and Lin, Rick and Liu, Zhongling and Padilla, Jake and Delimitrou, Christina},
title = {An Open-Source Benchmark Suite for Microservices and Their Hardware-Software Implications for Cloud \& Edge Systems},
year = {2019},
isbn = {9781450362405},
publisher = {Association for Computing Machinery},
address = {New York, NY, USA},
url = {https://doi.org/10.1145/3297858.3304013},
doi = {10.1145/3297858.3304013},
pages = {3–18},
numpages = {16},
location = {Providence, RI, USA},
series = {ASPLOS '19}
}

@inproceedings{daglis-asplos19-rpcvalet,
author = {Daglis, Alexandros and Sutherland, Mark and Falsafi, Babak},
title = {RPCValet: NI-Driven Tail-Aware Balancing of µs-Scale RPCs},
year = {2019},
isbn = {9781450362405},
publisher = {Association for Computing Machinery},
address = {New York, NY, USA},
url = {https://doi.org/10.1145/3297858.3304070},
doi = {10.1145/3297858.3304070},
booktitle = {Proceedings of the Twenty-Fourth International Conference on Architectural Support for Programming Languages and Operating Systems},
pages = {35–48},
numpages = {14},
location = {Providence, RI, USA},
series = {ASPLOS '19}
}

@INPROCEEDINGS{shuang-iiswc17-bigsmall,
  author={Chen, Shuang and GalOn, Shay and Delimitrou, Christina and Manne, Srilatha and Martínez, José F.},
  booktitle={2017 IEEE International Symposium on Workload Characterization (IISWC)}, 
  title={Workload characterization of interactive cloud services on big and small server platforms}, 
  year={2017},
  volume={},
  number={},
  pages={125-134},
  doi={10.1109/IISWC.2017.8167770}}

@ARTICLE{gan-cal18-archimpl,
  author={Gan, Yu and Delimitrou, Christina},
  journal={IEEE Computer Architecture Letters}, 
  title={The Architectural Implications of Cloud Microservices}, 
  year={2018},
  volume={17},
  number={2},
  pages={155-158},
  doi={10.1109/LCA.2018.2839189}}

@inproceedings{neugebauer-sigcomm18-pcieperf,
author = {Neugebauer, Rolf and Antichi, Gianni and Zazo, Jos\'{e} Fernando and Audzevich, Yury and L\'{o}pez-Buedo, Sergio and Moore, Andrew W.},
title = {Understanding PCIe performance for end host networking},
year = {2018},
isbn = {9781450355674},
publisher = {Association for Computing Machinery},
address = {New York, NY, USA},
url = {https://doi.org/10.1145/3230543.3230560},
doi = {10.1145/3230543.3230560},
booktitle = {Proceedings of the 2018 Conference of the ACM Special Interest Group on Data Communication},
pages = {327–341},
numpages = {15},
location = {Budapest, Hungary},
series = {SIGCOMM '18}
}

@INPROCEEDINGS{sutherland-isca20-nebula,
  author={Sutherland, Mark and Gupta, Siddharth and Falsafi, Babak and Marathe, Virendra and Pnevmatikatos, Dionisios and Daglis, Alexandros},
  booktitle={2020 ACM/IEEE 47th Annual International Symposium on Computer Architecture (ISCA)}, 
  title={The NEBULA RPC-Optimized Architecture}, 
  year={2020},
  volume={},
  number={},
  pages={199-212},
  doi={10.1109/ISCA45697.2020.00027}}

@inproceedings{pourhabibi-micor21-cerebros,
author = {Pourhabibi, Arash and Sutherland, Mark and Daglis, Alexandros and Falsafi, Babak},
title = {Cerebros: Evading the RPC Tax in Datacenters},
year = {2021},
isbn = {9781450385572},
publisher = {Association for Computing Machinery},
address = {New York, NY, USA},
url = {https://doi.org/10.1145/3466752.3480055},
doi = {10.1145/3466752.3480055},
booktitle = {MICRO-54: 54th Annual IEEE/ACM International Symposium on Microarchitecture},
pages = {407–420},
numpages = {14},
location = {Virtual Event, Greece},
series = {MICRO '21}
}

@article{wang-acm22-dcaperf,
author = {Wang, Minhu and Xu, Mingwei and Wu, Jianping},
title = {Understanding I/O Direct Cache Access Performance for End Host Networking},
year = {2022},
issue_date = {March 2022},
publisher = {Association for Computing Machinery},
address = {New York, NY, USA},
volume = {6},
number = {1},
url = {https://doi.org/10.1145/3508042},
doi = {10.1145/3508042},
journal = {Proc. ACM Meas. Anal. Comput. Syst.},
month = feb,
articleno = {22},
numpages = {37}
}

@inproceedings{lazarev-asplos21-dagger,
author = {Lazarev, Nikita and Xiang, Shaojie and Adit, Neil and Zhang, Zhiru and Delimitrou, Christina},
title = {Dagger: efficient and fast RPCs in cloud microservices with near-memory reconfigurable NICs},
year = {2021},
isbn = {9781450383172},
publisher = {Association for Computing Machinery},
address = {New York, NY, USA},
url = {https://doi.org/10.1145/3445814.3446696},
doi = {10.1145/3445814.3446696},
booktitle = {Proceedings of the 26th ACM International Conference on Architectural Support for Programming Languages and Operating Systems},
pages = {36–51},
numpages = {16},
location = {Virtual, USA},
series = {ASPLOS '21}
}

@INPROCEEDINGS{zhang-hpca25-rpcnic,
  author={Zhang, Jie and Huang, Hongjing and Chen, Xuzheng and Li, Xiang and Zhao, Jieru and Liu, Ming and Wang, Zeke},
  booktitle={2025 IEEE International Symposium on High Performance Computer Architecture (HPCA)}, 
  title={RpcNIC: Enabling Efficient Datacenter RPC Offloading on PCIe-attached SmartNICs}, 
  year={2025},
  volume={},
  number={},
  pages={1379-1394},
  doi={10.1109/HPCA61900.2025.00104}}

@INPROCEEDINGS{ueda-iiswc16-workloadcharacterization,
  author={Ueda, Takanori and Nakaike, Takuya and Ohara, Moriyoshi},
  booktitle={2016 IEEE International Symposium on Workload Characterization (IISWC)}, 
  title={Workload characterization for microservices}, 
  year={2016},
  volume={},
  number={},
  pages={1-10},
  doi={10.1109/IISWC.2016.7581269}}

@inproceedings{cai-sigcomm21-understandinghoststackoverhead,
author = {Cai, Qizhe and Chaudhary, Shubham and Vuppalapati, Midhul and Hwang, Jaehyun and Agarwal, Rachit},
title = {Understanding host network stack overheads},
year = {2021},
isbn = {9781450383837},
publisher = {Association for Computing Machinery},
address = {New York, NY, USA},
url = {https://doi.org/10.1145/3452296.3472888},
doi = {10.1145/3452296.3472888},
booktitle = {Proceedings of the 2021 ACM SIGCOMM 2021 Conference},
pages = {65–77},
numpages = {13},
location = {Virtual Event, USA},
series = {SIGCOMM '21}
}

@inproceedings {skiadopoulos-osdi24-flexiblehostnetworking,
author = {Athinagoras Skiadopoulos and Zhiqiang Xie and Mark Zhao and Qizhe Cai and Saksham Agarwal and Jacob Adelmann and David Ahern and Carlo Contavalli and Michael Goldflam and Vitaly Mayatskikh and Raghu Raja and Daniel Walton and Rachit Agarwal and Shrijeet Mukherjee and Christos Kozyrakis},
title = {High-throughput and Flexible Host Networking for Accelerated Computing},
booktitle = {18th USENIX Symposium on Operating Systems Design and Implementation (OSDI 24)},
year = {2024},
isbn = {978-1-939133-40-3},
address = {Santa Clara, CA},
pages = {405--423},
url = {https://www.usenix.org/conference/osdi24/presentation/skiadopoulos},
publisher = {USENIX Association},
month = jul
}

@inproceedings{wolnikowski-hotos21-zerializer,
  title={Zerializer: Towards zero-copy serialization},
  author={Wolnikowski, Adam and Ibanez, Stephen and Stone, Jonathan and Kim, Changhoon and Manohar, Rajit and Soul{\'e}, Robert},
  booktitle={Proceedings of the Workshop on Hot Topics in Operating Systems},
  pages={206--212},
  year={2021}
}

@INPROCEEDINGS{shao-micro16-gem5aladdin,
  author={Shao, Yakun Sophia and Xi, Sam Likun and Srinivasan, Vijayalakshmi and Wei, Gu-Yeon and Brooks, David},
  booktitle={2016 49th Annual IEEE/ACM International Symposium on Microarchitecture (MICRO)}, 
  title={Co-designing accelerators and SoC interfaces using gem5-Aladdin}, 
  year={2016},
  volume={},
  number={},
  pages={1-12},
  doi={10.1109/MICRO.2016.7783751}
}

@article{izmirli-cal25-halis,
  title={Halis: A Hardware-Software Co-designed Near-Cache Accelerator for Graph Pattern Mining},
  author={{\.I}zmirli, {\c{S}}evval and Pavon, Julian and Valdivieso, Ivan Vargas and Ayde{\u{g}}er, Bet{\"u}l and Yal{\c{c}}{\i}nkaya, Kerem and Cristal, Adrian and Ergin, O{\u{g}}uz and {\"U}nsal, Osman},
  journal={IEEE Computer Architecture Letters},
  year={2025},
  publisher={IEEE}
}

@inproceedings{umeike-ispass24-userspace,
  title={Userspace Networking in gem5},
  author={Umeike, Johnson and Agarwal, Siddharth and Lazarev, Nikita and Alian, Mohammad},
  booktitle={2024 IEEE International Symposium on Performance Analysis of Systems and Software (ISPASS)},
  pages={179--191},
  year={2024},
  organization={IEEE}
}

@article{denzler-ieee23-casper,
  title={Casper: Accelerating stencil computations using near-cache processing},
  author={Denzler, Alain and Oliveira, Geraldo F and Hajinazar, Nastaran and Bera, Rahul and Singh, Gagandeep and G{\'o}mez-Luna, Juan and Mutlu, Onur},
  journal={IEEE Access},
  volume={11},
  pages={22136--22154},
  year={2023},
  publisher={IEEE}
}

@inproceedings{wang-hpca22-nearstream,
  title={Near-stream computing: General and transparent near-cache acceleration},
  author={Wang, Zhengrong and Weng, Jian and Liu, Sihao and Nowatzki, Tony},
  booktitle={2022 IEEE International Symposium on High-Performance Computer Architecture (HPCA)},
  pages={331--345},
  year={2022},
  organization={IEEE}
}

@inproceedings{nori-isca21-reduct,
  title={REDUCT: Keep it close, keep it cool!: Efficient scaling of DNN inference on multi-core CPUs with near-cache compute},
  author={Nori, Anant V and Bera, Rahul and Balachandran, Shankar and Rakshit, Joydeep and Omer, Om J and Abuhatzera, Avishaii and Kuttanna, Belliappa and Subramoney, Sreenivas},
  booktitle={2021 ACM/IEEE 48th Annual International Symposium on Computer Architecture (ISCA)},
  pages={167--180},
  year={2021},
  organization={IEEE}
}

@article{binkert-sigarch11-gem5,
author = {Binkert, Nathan and Beckmann, Bradford and Black, Gabriel and Reinhardt, Steven K. and Saidi, Ali and Basu, Arkaprava and Hestness, Joel and Hower, Derek R. and Krishna, Tushar and Sardashti, Somayeh and Sen, Rathijit and Sewell, Korey and Shoaib, Muhammad and Vaish, Nilay and Hill, Mark D. and Wood, David A.},
title = {The gem5 simulator},
year = {2011},
issue_date = {May 2011},
publisher = {Association for Computing Machinery},
address = {New York, NY, USA},
volume = {39},
number = {2},
issn = {0163-5964},
url = {https://doi.org/10.1145/2024716.2024718},
doi = {10.1145/2024716.2024718},
journal = {SIGARCH Comput. Archit. News},
month = aug,
pages = {1–7},
numpages = {7}
}

@techreport{nvidia_infiniband,
 author = {NVIDIA},
 title = {InfiniBand Technology Overview},
 institution = {NVIDIA Corporation},
 year = {2025},
 type = {White Paper},
 url = {https://network.nvidia.com/pdf/whitepapers/WP_InfiniBand_Technology_Overview.pdf},
 note = {Accessed: 2025-07-26}
}

@misc{nvidia_roce,
 author = {{NVIDIA}},
 title = {RDMA over Converged Ethernet (RoCE)},
 year = {2025},
 url ={https://docs.nvidia.com/networking/display/mlnxofedv23070512/rdma+over+converged+ethernet+(roce)},
 note = {NVIDIA Networking Documentation, Accessed: 2025-07-26}
}

@inproceedings{yuan-24-intelaia,
  title={Intel accelerators ecosystem: An soc-oriented perspective: Industry product},
  author={Yuan, Yifan and Wang, Ren and Ranganathan, Narayan and Rao, Nikhil and Kumar, Sanjay and Lantz, Philip and Sanjeepan, Vivekananthan and Cabrera, Jorge and Kwatra, Atul and Sankaran, Rajesh and others},
  booktitle={2024 ACM/IEEE 51st Annual International Symposium on Computer Architecture (ISCA)},
  pages={848--862},
  year={2024},
  organization={IEEE}
}

@misc{arm_cp14_cp15,
 author = {Arm},
 title = {{CP14 and CP15 system control registers}},
 year = {2025},
 url = {https://developer.arm.com/documentation/ddi0406/cb/System-Level-Architecture/The-System-Level-Programmers--Model/Coprocessors-and-system-control/CP14-and-CP15-system-control-registers?lang=en},
 note = {Arm Developer Documentation, Accessed: 2025-07-26}
}

@inproceedings{sriraman-iiswc18-microsuite,
  title={$\mu$ suite: a benchmark suite for microservices},
  author={Sriraman, Akshitha and Wenisch, Thomas F},
  booktitle={2018 ieee international symposium on workload characterization (iiswc)},
  pages={1--12},
  year={2018},
  organization={IEEE}
}

@inproceedings{pourhabibi2020optimus,
  title={Optimus prime: Accelerating data transformation in servers},
  author={Pourhabibi, Arash and Gupta, Siddharth and Kassir, Hussein and Sutherland, Mark and Tian, Zilu and Drumond, Mario Paulo and Falsafi, Babak and Koch, Christoph},
  booktitle={Proceedings of the Twenty-Fifth International Conference on Architectural Support for Programming Languages and Operating Systems},
  pages={1203--1216},
  year={2020}
}

@inproceedings{karandikar-micro21-protoacc,
  title={A hardware accelerator for protocol buffers},
  author={Karandikar, Sagar and Leary, Chris and Kennelly, Chris and Zhao, Jerry and Parimi, Dinesh and Nikolic, Borivoje and Asanovic, Krste and Ranganathan, Parthasarathy},
  booktitle={MICRO-54: 54th Annual IEEE/ACM International Symposium on Microarchitecture},
  pages={462--478},
  year={2021}
}

@inproceedings{marty2019snap,
  title={Snap: A microkernel approach to host networking},
  author={Marty, Michael and de Kruijf, Marc and Adriaens, Jacob and Alfeld, Christopher and Bauer, Sean and Contavalli, Carlo and Dalton, Michael and Dukkipati, Nandita and Evans, William C and Gribble, Steve and others},
  booktitle={Proceedings of the 27th ACM Symposium on Operating Systems Principles},
  pages={399--413},
  year={2019}
}

@inproceedings{belay2014ix,
  title={$\{$IX$\}$: a protected dataplane operating system for high throughput and low latency},
  author={Belay, Adam and Prekas, George and Klimovic, Ana and Grossman, Samuel and Kozyrakis, Christos and Bugnion, Edouard},
  booktitle={11th USENIX Symposium on Operating Systems Design and Implementation (OSDI 14)},
  pages={49--65},
  year={2014}
}

@inproceedings{prekas2017zygos,
  title={Zygos: Achieving low tail latency for microsecond-scale networked tasks},
  author={Prekas, George and Kogias, Marios and Bugnion, Edouard},
  booktitle={Proceedings of the 26th Symposium on Operating Systems Principles},
  pages={325--341},
  year={2017}
}

@inproceedings{ousterhout2019shenango,
  title={Shenango: Achieving high $\{$CPU$\}$ efficiency for latency-sensitive datacenter workloads},
  author={Ousterhout, Amy and Fried, Joshua and Behrens, Jonathan and Belay, Adam and Balakrishnan, Hari},
  booktitle={16th USENIX Symposium on Networked Systems Design and Implementation (NSDI 19)},
  pages={361--378},
  year={2019}
}

@techreport{nvidia_rdma_routed,
 author = {{NVIDIA}},
 title = {{Benefits of RDMA over Routed Fabrics}},
 institution = {NVIDIA Corporation},
 year = {2025},
 type = {Solution Brief},
 url = {https://network.nvidia.com/pdf/solutions/benefits-of-RDMA-over-routed-fabrics.pdf},
 note = {Accessed: 2025-07-26}
}

@inproceedings{dragojevic2014farm,
  title={$\{$FaRM$\}$: Fast remote memory},
  author={Dragojevi{\'c}, Aleksandar and Narayanan, Dushyanth and Castro, Miguel and Hodson, Orion},
  booktitle={11th USENIX Symposium on Networked Systems Design and Implementation (NSDI 14)},
  pages={401--414},
  year={2014}
}

@inproceedings {kalia2016fasst,
author = {Anuj Kalia and Michael Kaminsky and David G. Andersen},
title = {{FaSST}: Fast, Scalable and Simple Distributed Transactions with Two-Sided (RDMA) Datagram {RPCs}},
booktitle = {12th USENIX Symposium on Operating Systems Design and Implementation (OSDI 16)},
year = {2016},
isbn = {978-1-931971-33-1},
address = {Savannah, GA},
pages = {185--201},
url = {https://www.usenix.org/conference/osdi16/technical-sessions/presentation/kalia},
publisher = {USENIX Association},
month = nov
}

@inproceedings{kalia2019datacenter,
  title={Datacenter $\{$RPCs$\}$ can be general and fast},
  author={Kalia, Anuj and Kaminsky, Michael and Andersen, David},
  booktitle={16th USENIX Symposium on Networked Systems Design and Implementation (NSDI 19)},
  pages={1--16},
  year={2019}
}

@misc{corba_wiki,
 author = {Wikipedia},
 title = {Common Object Request Broker Architecture},
 year = {2025},
 url = {https://en.wikipedia.org/wiki/Common_Object_Request_Broker_Architecture},
 note = {Accessed: 2025-07-26}
}

@misc{sunrpc_wiki,
 author = {{Wikipedia}},
 title = {{Sun RPC}},
 year = {2025},
 url = {https://en.wikipedia.org/wiki/Sun_RPC},
 note = {Accessed: 2025-07-26}
}

@inproceedings{zhang2019muqsim,
  title={$\mu$qsim: Enabling accurate and scalable simulation for interactive microservices},
  author={Zhang, Yanqi and Gan, Yu and Delimitrou, Christina},
  booktitle={2019 IEEE International Symposium on Performance Analysis of Systems and Software (ISPASS)},
  pages={212--222},
  year={2019},
  organization={IEEE}
}

@inproceedings{zhou2018benchmarking,
  title={Benchmarking microservice systems for software engineering research},
  author={Zhou, Xiang and Peng, Xin and Xie, Tao and Sun, Jun and Xu, Chenjie and Ji, Chao and Zhao, Wenyun},
  booktitle={Proceedings of the 40th International Conference on Software Engineering: Companion Proceeedings},
  pages={323--324},
  year={2018}
}

@inproceedings{zhao2022altocumulus,
  title={Altocumulus: Scalable scheduling for nanosecond-scale remote procedure calls},
  author={Zhao, Jiechen and Uwizeyimana, Iris and Ganesan, Karthik and Jeffrey, Mark C and Jerger, Natalie Enright},
  booktitle={2022 55th IEEE/ACM International Symposium on Microarchitecture (MICRO)},
  pages={423--440},
  year={2022},
  organization={IEEE}
}

@misc{capnproto2025,
 author = {{Cap'n Proto}},
 title = {{Cap'n Proto - Infinitely Fast Serialization}},
 year = {2025},
 url = {https://capnproto.org/},
 note = {Accessed: 2025-07-26}
}

@inproceedings{khairy2022simr,
  title={Simr: Single instruction multiple request processing for energy-efficient data center microservices},
  author={Khairy, Mahmoud and Alawneh, Ahmad and Barnes, Aaron and Rogers, Timothy G},
  booktitle={2022 55th IEEE/ACM International Symposium on Microarchitecture (MICRO)},
  pages={441--463},
  year={2022},
  organization={IEEE}
}

@inproceedings{jeong2014mtcp,
  title={$\{$mTCP$\}$: a highly scalable user-level $\{$TCP$\}$ stack for multicore systems},
  author={Jeong, EunYoung and Wood, Shinae and Jamshed, Muhammad and Jeong, Haewon and Ihm, Sunghwan and Han, Dongsu and Park, KyoungSoo},
  booktitle={11th USENIX Symposium on Networked Systems Design and Implementation (NSDI 14)},
  pages={489--502},
  year={2014}
}

@inproceedings{fried2020caladan,
  title={Caladan: Mitigating interference at microsecond timescales},
  author={Fried, Joshua and Ruan, Zhenyuan and Ousterhout, Amy and Belay, Adam},
  booktitle={14th USENIX Symposium on Operating Systems Design and Implementation (OSDI 20)},
  pages={281--297},
  year={2020}
}

@inproceedings{eran2019nica,
  title={$\{$NICA$\}$: An infrastructure for inline acceleration of network applications},
  author={Eran, Haggai and Zeno, Lior and Tork, Maroun and Malka, Gabi and Silberstein, Mark},
  booktitle={2019 USENIX Annual Technical Conference (USENIX ATC 19)},
  pages={345--362},
  year={2019}
}

@inproceedings{firestone2018azure,
  title={Azure accelerated networking:$\{$SmartNICs$\}$ in the public cloud},
  author={Firestone, Daniel and Putnam, Andrew and Mundkur, Sambhrama and Chiou, Derek and Dabagh, Alireza and Andrewartha, Mike and Angepat, Hari and Bhanu, Vivek and Caulfield, Adrian and Chung, Eric and others},
  booktitle={15th USENIX Symposium on Networked Systems Design and Implementation (NSDI 18)},
  pages={51--66},
  year={2018}
}

@inproceedings{alian2019netdimm,
  title={Netdimm: Low-latency near-memory network interface architecture},
  author={Alian, Mohammad and Kim, Nam Sung},
  booktitle={Proceedings of the 52nd Annual IEEE/ACM International Symposium on Microarchitecture},
  pages={699--711},
  year={2019}
}

@article{ghose2019processing,
  title={Processing-in-memory: A workload-driven perspective},
  author={Ghose, Saugata and Boroumand, Amirali and Kim, Jeremie S and G{\'o}mez-Luna, Juan and Mutlu, Onur},
  journal={IBM Journal of Research and Development},
  volume={63},
  number={6},
  pages={3--1},
  year={2019},
  publisher={IBM}
}

@inproceedings{esmaeilzadeh2011dark,
  title={Dark silicon and the end of multicore scaling},
  author={Esmaeilzadeh, Hadi and Blem, Emily and St. Amant, Renee and Sankaralingam, Karthikeyan and Burger, Doug},
  booktitle={Proceedings of the 38th annual international symposium on Computer architecture},
  pages={365--376},
  year={2011}
}

@inproceedings{caulfield2016cloud,
  title={A cloud-scale acceleration architecture},
  author={Caulfield, Adrian M and Chung, Eric S and Putnam, Andrew and Angepat, Hari and Fowers, Jeremy and Haselman, Michael and Heil, Stephen and Humphrey, Matt and Kaur, Puneet and Kim, Joo-Young and others},
  booktitle={2016 49th Annual IEEE/ACM international symposium on microarchitecture (MICRO)},
  pages={1--13},
  year={2016},
  organization={IEEE}
}

@inproceedings{seyedroudbari2023turbo,
  title={Turbo: Smartnic-enabled dynamic load balancing of $\mu$s-scale rpcs},
  author={Seyedroudbari, Hamed and Vanavasam, Srikar and Daglis, Alexandros},
  booktitle={2023 IEEE International Symposium on High-Performance Computer Architecture (HPCA)},
  pages={1045--1058},
  year={2023},
  organization={IEEE}
}

@inproceedings{huang2024hal,
  title={Hal: Hardware-assisted load balancing for energy-efficient snic-host cooperative computing},
  author={Huang, Jinghan and Lou, Jiaqi and Vanavasam, Srikar and Kong, Xinhao and Ji, Houxiang and Jeong, Ipoom and Zhuo, Danyang and Lee, Eun Kyung and Kim, Nam Sung},
  booktitle={2024 ACM/IEEE 51st Annual International Symposium on Computer Architecture (ISCA)},
  pages={613--627},
  year={2024},
  organization={IEEE}
}

@inproceedings{mirhosseini2019enhancing,
  title={Enhancing server efficiency in the face of killer microseconds},
  author={Mirhosseini, Amirhossein and Sriraman, Akshitha and Wenisch, Thomas F},
  booktitle={2019 IEEE International Symposium on High Performance Computer Architecture (HPCA)},
  pages={185--198},
  year={2019},
  organization={IEEE}
}

@inproceedings{kim2017toward,
  title={Toward standardized near-data processing with unrestricted data placement for GPUs},
  author={Kim, Gwangsun and Chatterjee, Niladrish and O'Connor, Mike and Hsieh, Kevin},
  booktitle={Proceedings of the International Conference for High Performance Computing, Networking, Storage and Analysis},
  pages={1--12},
  year={2017}
}

@inproceedings{singh2018review,
  title={A review of near-memory computing architectures: Opportunities and challenges},
  author={Singh, Gagandeep and Chelini, Lorenzo and Corda, Stefano and Awan, Ahsan Javed and Stuijk, Sander and Jordans, Roel and Corporaal, Henk and Boonstra, Albert-Jan},
  booktitle={2018 21st Euromicro Conference on Digital System Design (DSD)},
  pages={608--617},
  year={2018},
  organization={IEEE}
}

@inproceedings{lee2021hardware,
  title={Hardware architecture and software stack for PIM based on commercial DRAM technology: Industrial product},
  author={Lee, Sukhan and Kang, Shin-haeng and Lee, Jaehoon and Kim, Hyeonsu and Lee, Eojin and Seo, Seungwoo and Yoon, Hosang and Lee, Seungwon and Lim, Kyounghwan and Shin, Hyunsung and others},
  booktitle={2021 ACM/IEEE 48th Annual International Symposium on Computer Architecture (ISCA)},
  pages={43--56},
  year={2021},
  organization={IEEE}
}

@incollection{mutlu2022modern,
  title={A modern primer on processing in memory},
  author={Mutlu, Onur and Ghose, Saugata and G{\'o}mez-Luna, Juan and Ausavarungnirun, Rachata},
  booktitle={Emerging computing: from devices to systems: looking beyond Moore and Von Neumann},
  pages={171--243},
  year={2022},
  publisher={Springer}
}

@inproceedings{heirman2018near,
  title={Near-side prefetch throttling: Adaptive prefetching for high-performance many-core processors},
  author={Heirman, Wim and Bois, Kristof Du and Vandriessche, Yves and Eyerman, Stijn and Hur, Ibrahim},
  booktitle={Proceedings of the 27th international conference on parallel architectures and compilation techniques},
  pages={1--11},
  year={2018}
}

@inproceedings{deb2019performance,
  title={Performance enhancement of caches in TCMPs using near vicinity prefetcher},
  author={Deb, Dipika and Jose, John and Palesi, Maurizio},
  booktitle={2019 32nd International Conference on VLSI Design and 2019 18th International Conference on Embedded Systems (VLSID)},
  pages={13--18},
  year={2019},
  organization={IEEE Computer Society}
}

@inproceedings{boroumand2019conda,
  title={CoNDA: Efficient cache coherence support for near-data accelerators},
  author={Boroumand, Amirali and Ghose, Saugata and Patel, Minesh and Hassan, Hasan and Lucia, Brandon and Ausavarungnirun, Rachata and Hsieh, Kevin and Hajinazar, Nastaran and Malladi, Krishna T and Zheng, Hongzhong and others},
  booktitle={Proceedings of the 46th International Symposium on Computer Architecture},
  pages={629--642},
  year={2019}
}

@article{valls2015ps,
  title={Ps directory: A scalable multilevel directory cache for cmps},
  author={Valls, Joan J and Ros, Alberto and Sahuquillo, Julio and G{\'o}mez, Mar{\'\i}a E},
  journal={The Journal of Supercomputing},
  volume={71},
  number={8},
  pages={2847--2876},
  year={2015},
  publisher={Springer}
}

@article{peter2015arrakis,
  title={Arrakis: The operating system is the control plane},
  author={Peter, Simon and Li, Jialin and Zhang, Irene and Ports, Dan RK and Woos, Doug and Krishnamurthy, Arvind and Anderson, Thomas and Roscoe, Timothy},
  journal={ACM Transactions on Computer Systems (TOCS)},
  volume={33},
  number={4},
  pages={1--30},
  year={2015},
  publisher={ACM New York, NY, USA}
}

@inproceedings{li2015architecting,
  title={Architecting to achieve a billion requests per second throughput on a single key-value store server platform},
  author={Li, Sheng and Lim, Hyeontaek and Lee, Victor W and Ahn, Jung Ho and Kalia, Anuj and Kaminsky, Michael and Andersen, David G and Seongil, O and Lee, Sukhan and Dubey, Pradeep},
  booktitle={Proceedings of the 42nd Annual International Symposium on Computer Architecture},
  pages={476--488},
  year={2015}
}

@inproceedings{farshin2020reexamining,
  title={Reexamining direct cache access to optimize $\{$I/O$\}$ intensive applications for multi-hundred-gigabit networks},
  author={Farshin, Alireza and Roozbeh, Amir and Maguire Jr, Gerald Q and Kosti{\'c}, Dejan},
  booktitle={2020 USENIX Annual Technical Conference (USENIX ATC 20)},
  pages={673--689},
  year={2020}
}

@ARTICLE{Lee2022SMIV,
  author={Lee, Sae Kyu and Whatmough, Paul N. and Donato, Marco and Ko, Glenn G. and Brooks, David and Wei, Gu-Yeon},
  journal={IEEE Journal of Solid-State Circuits}, 
  title={SMIV: A 16-nm 25-mm² SoC for IoT With Arm Cortex-A53, eFPGA, and Coherent Accelerators}, 
  year={2022},
  volume={57},
  number={2},
  pages={639-650},
  doi={10.1109/JSSC.2021.3115466}}

@inproceedings{kuper2024quantitative,
  title={A quantitative analysis and guidelines of data streaming accelerator in modern intel xeon scalable processors},
  author={Kuper, Reese and Jeong, Ipoom and Yuan, Yifan and Wang, Ren and Ranganathan, Narayan and Rao, Nikhil and Hu, Jiayu and Kumar, Sanjay and Lantz, Philip and Kim, Nam Sung},
  booktitle={Proceedings of the 29th ACM International Conference on Architectural Support for Programming Languages and Operating Systems, Volume 2},
  pages={37--54},
  year={2024}
}

@misc{intel_dsa,
  author = {Intel Corporation},
  title = {{Intel Data Streaming Accelerator}},
  year = {2025},
  url = {https://www.intel.com/content/www/us/en/products/docs/accelerator-engines/data-streaming-accelerator.html},
  note = {Accessed: 2025-07-26}
}

@inproceedings{Luo2021AlibabaMicroservice,
author = {Luo, Shutian and Xu, Huanle and Lu, Chengzhi and Ye, Kejiang and Xu, Guoyao and Zhang, Liping and Ding, Yu and He, Jian and Xu, Chengzhong},
title = {Characterizing Microservice Dependency and Performance: Alibaba Trace Analysis},
year = {2021},
isbn = {9781450386388},
publisher = {Association for Computing Machinery},
address = {New York, NY, USA},
url = {https://doi.org/10.1145/3472883.3487003},
doi = {10.1145/3472883.3487003},
booktitle = {Proceedings of the ACM Symposium on Cloud Computing},
pages = {412–426},
numpages = {15},
location = {Seattle, WA, USA},
series = {SoCC '21}
}

@inproceedings{2021CarreiraFromWarmToHotStart,
author = {Carreira, Joao and Kohli, Sumer and Bruno, Rodrigo and Fonseca, Pedro},
title = {From warm to hot starts: leveraging runtimes for the serverless era},
year = {2021},
isbn = {9781450384384},
publisher = {Association for Computing Machinery},
address = {New York, NY, USA},
url = {https://doi.org/10.1145/3458336.3465305},
doi = {10.1145/3458336.3465305},
booktitle = {Proceedings of the Workshop on Hot Topics in Operating Systems},
pages = {58–64},
numpages = {7},
location = {Ann Arbor, Michigan},
series = {HotOS '21}
}

@inproceedings{2025ColdStarts,
author = {Joosen, Artjom and Hassan, Ahmed and Asenov, Martin and Singh, Rajkarn and Darlow, Luke and Wang, Jianfeng and Deng, Qiwen and Barker, Adam},
title = {Serverless Cold Starts and Where to Find Them},
year = {2025},
isbn = {9798400711961},
publisher = {Association for Computing Machinery},
address = {New York, NY, USA},
url = {https://doi.org/10.1145/3689031.3696073},
doi = {10.1145/3689031.3696073},
pages = {938–953},
numpages = {16},
location = {Rotterdam, Netherlands},
series = {EuroSys '25}
}

@INPROCEEDINGS{Zynq7000,
  author={Rajagopalan, Vidya and Boppana, Vamsi and Dutta, Sandeep and Taylor, Brad and Wittig, Ralph},
  booktitle={2011 IEEE Hot Chips 23 Symposium (HCS)}, 
  title={Xilinx Zynq-7000 EPP: An extensible processing platform family}, 
  year={2011},
  volume={},
  number={},
  pages={1-24},
  doi={10.1109/HOTCHIPS.2011.7477495}}

@misc{quicklogic2021efpga,
  author       = {{QuickLogic Corporation}},
  title        = {{QuickLogic Announces Contract for High-Density eFPGA Hard IP}},
  year         = {2021},
  howpublished = {\url{https://ir.quicklogic.com/press-releases/detail/766/quicklogic-announces-contract-for-high-density-efpga-hard}},
  note         = {Accessed: 2026-04-07}
}

@inproceedings{2022TLBshootdown,
author = {Amit, Nadav and Tai, Amy and Wei, Michael},
title = {Don't shoot down TLB shootdowns!},
year = {2020},
isbn = {9781450368827},
publisher = {Association for Computing Machinery},
address = {New York, NY, USA},
url = {https://doi.org/10.1145/3342195.3387518},
doi = {10.1145/3342195.3387518},
booktitle = {Proceedings of the Fifteenth European Conference on Computer Systems},
articleno = {35},
numpages = {14},
location = {Heraklion, Greece},
series = {EuroSys '20}
}

@INPROCEEDINGS{ultra_dense_sram,
  author={Nautiyal, Vivek and Singla, Gaurav and Gupta, Lalit and Dwivedi, Sagar and Kinkade, Martin},
  booktitle={2017 30th IEEE International System-on-Chip Conference (SOCC)}, 
  title={An ultra high density pseudo dual-port SRAM in 16nm FINFET process for graphics processors}, 
  year={2017},
  volume={},
  number={},
  pages={12-17},
  doi={10.1109/SOCC.2017.8225996}}

@misc{vu19p_die_size,
  title = {Xilinx Introduces World’s Largest FPGA With 35 Billion Transistors
},
  author={Tom's Hardware},
  howpublished = {\url{https://www.tomshardware.com/news/xilinx-world-largest-fpga,40212.html}},
  note = {Accessed: 2026-6-15}
}

@misc{vu19p,
  title = {{U}ltrascale-plus {FPGA} product selection guide
},
  author={AMD},
  howpublished = {\url{https://docs.amd.com/v/u/en-US/ultrascale-plus-fpga-product-selection-guide}},
  note = {Accessed: 2026-6-15}
}

@misc{Adicsys,
  title = {Synthesizable Programmable Core
},
  author={Adicsys},
  howpublished = {\url{http://www.adicsys.com/uploads/pdf/spc_datasheet.pdf}},
  note = {Accessed: 2026-6-15}
}

@misc{flex-logix,
  title = {EFLX® Embedded FPGA
},
  author={Flex logix},
  howpublished = {\url{https://assets.flex-logix.com/resources/2022\%2006\%20EFLX\%204-page\%20Overview\%20TGF-eFPGA.pdf}},
  note = {Accessed: 2026-6-15}
}

@inproceedings{kuon2006,
  author    = {Ian Kuon and Jonathan Rose},
  title     = {Measuring the Gap Between {FPGAs} and {ASICs}},
  booktitle = {Proceedings of the ACM/SIGDA International Symposium
               on Field Programmable Gate Arrays},
  year      = {2006},
  pages     = {21--30},
  doi       = {10.1145/1117201.1117205}
}

@misc{icap,
  author       = {{AMD/Xilinx}},
  title        = {{Vivado Design Suite User Guide: Partial Reconfiguration (UG909)}},
  howpublished = {\url{https://docs.amd.com/r/en-US/ug909-vivado-partial-reconfiguration}},
  note         = {Accessed: 2026-6-15}
}

@inproceedings{mcpat,
author = {Li, Sheng and Ahn, Jung Ho and Strong, Richard D. and Brockman, Jay B. and Tullsen, Dean M. and Jouppi, Norman P.},
title = {{McPAT}: An Integrated Power, Area, and Timing Modeling Framework for Multicore and Manycore Architectures},
booktitle = {Proceedings of the 42nd Annual IEEE/ACM International Symposium on Microarchitecture},
pages = {469--480},
year = {2009},
series = {MICRO 42}
}

@misc{cacti,
  author       = {Shyamkumar Thoziyoor and Naveen Muralimanohar and Jung Ho Ahn and Norman P. Jouppi},
  title        = {{CACTI} 6.5},
  howpublished = {\url{https://github.com/HewlettPackard/cacti}},
  year         = {2008},
  note         = {Accessed: 2026-6-15}
}

@inproceedings{novakovic2014sonuma,
  author    = {Novakovic, Stanko and Daglis, Alexandros and Bugnion, Edouard and Falsafi, Babak and Grot, Boris},
  title     = {Scale-out {NUMA}},
  booktitle = {Proceedings of the 19th International Conference on Architectural Support for Programming Languages and Operating Systems (ASPLOS)},
  year      = {2014},
  pages     = {3--18},
  doi       = {10.1145/2541940.2541965}
}

@inproceedings{daglis2015manycore,
  author    = {Daglis, Alexandros and Novakovi\'{c}, Stanko and Bugnion, Edouard and Falsafi, Babak and Grot, Boris},
  title     = {Manycore Network Interfaces for In-Memory Rack-Scale Computing},
  booktitle = {Proceedings of the 42nd Annual International Symposium on Computer Architecture (ISCA)},
  year      = {2015},
  pages     = {567--579},
  doi       = {10.1145/2749469.2750415}
}

@inproceedings{daglis2019rpcvalet,
  author    = {Daglis, Alexandros and Sutherland, Mark and Falsafi, Babak},
  title     = {{RPCValet}: {NI}-Driven Tail-Aware Balancing of $\mu$s-Scale {RPCs}},
  booktitle = {Proceedings of the 24th International Conference on Architectural Support for Programming Languages and Operating Systems (ASPLOS)},
  year      = {2019},
  pages     = {35--48},
  doi       = {10.1145/3297858.3304070}
}

@manual{arm_amba_chi,
  author = {{Arm Limited}},
  title  = {AMBA 5 CHI Architecture Specification},
  year   = {2025},
  note   = {Arm IHI 0050, Issue H, \S B1.6},
  url    = {https://developer.arm.com/documentation/ihi0050/latest/}
}

@book{sparc_tso,
  editor    = {Weaver, David L. and Germond, Tom},
  title     = {The {SPARC} Architecture Manual, Version 9},
  publisher = {SPARC International, Inc., Prentice Hall},
  year      = {1994},
  isbn      = {0-13-825001-4},
  note      = {Ch.~8 (Memory Models) and Appendix~D.6 (Total Store Order)}
}

@misc{apple_tso,
  author       = {{Apple Inc.}},
  title        = {Accelerating the Performance of Rosetta},
  howpublished = {\url{https://developer.apple.com/documentation/virtualization/accelerating-the-performance-of-rosetta}},
  note         = {Total Store Ordering (TSO) memory model on Apple silicon; accessed 2026-07-27}
}

@article{x86_tso,
author = {Sewell, Peter and Sarkar, Susmit and Owens, Scott and Nardelli, Francesco Zappa and Myreen, Magnus O.},
title = {x86-TSO: a rigorous and usable programmer's model for x86 multiprocessors},
year = {2010},
issue_date = {July 2010},
publisher = {Association for Computing Machinery},
address = {New York, NY, USA},
volume = {53},
number = {7},
issn = {0001-0782},
url = {https://doi.org/10.1145/1785414.1785443},
doi = {10.1145/1785414.1785443},
journal = {Commun. ACM},
month = jul,
pages = {89–97},
numpages = {9}
}

@article{Ex86_tso,
  author  = {Raad, Azalea and Maranget, Luc and Vafeiadis, Viktor},
  title   = {Extending {Intel-x86} Consistency and Persistency: Formalising the Semantics of {Intel-x86} Memory Types and Non-Temporal Stores},
  journal = {Proc. ACM Program. Lang.},
  volume  = {6},
  number  = {POPL},
  articleno = {22},
  year    = {2022},
  doi     = {10.1145/3498683}
}

@misc{riscv_ztso,
  author       = {{RISC-V International}},
  title        = {The {RISC-V} Instruction Set Manual, Volume I: Unprivileged ISA, Chapter ``Ztso'' Extension for Total Store Ordering, Version 1.0},
  year         = {2024},
  howpublished = {\url{https://docs.riscv.org/reference/isa/unpriv/ztso-st-ext.html}},
  note         = {Accessed 2026-07-27}
}

@manual{amd-apm-vol2,
  author = {{Advanced Micro Devices}},
  title  = {{AMD64} Architecture Programmer's Manual, Volume 2: System Programming},
  year   = {2026},
  note   = {Rev. 3.44, \S 7.2, Multiprocessor Memory Access Ordering}
}

@manual{intel-sdm-vol3,
  author = {{Intel Corporation}},
  title  = {{Intel}\textregistered{} 64 and {IA-32} Architectures Software Developer's Manual, Volume 3A: System Programming Guide, Part 1},
  year   = {2026},
  month  = jun,
  note   = {Order Number 253668-092US, \S~11.2.2 (Memory Ordering) and \S~14.3 (Methods of Caching)}
}

@manual{arm_arm_ddi0487,
  author = {{Arm Limited}},
  title  = {Arm Architecture Reference Manual for A-profile Architecture},
  year   = {2026},
  note   = {Arm DDI 0487M.c, \S~B2.6.10 (Store-Release) and \S~C6.2 (STLR)}
}

@inproceedings{gharachorloo1990release,
  author    = {Gharachorloo, Kourosh and Lenoski, Daniel and Laudon, James and Gibbons, Phillip and Gupta, Anoop and Hennessy, John},
  title     = {Memory Consistency and Event Ordering in Scalable Shared-Memory Multiprocessors},
  booktitle = {Proceedings of the 17th Annual International Symposium on Computer Architecture (ISCA)},
  year      = {1990}, pages = {15--26}, doi = {10.1109/ISCA.1990.134503}
}

@manual{riscv_unpriv,
  author = {{RISC-V International}},
  title  = {The {RISC-V} Instruction Set Manual, Volume I: Unprivileged Architecture},
  year   = {2024},
  note   = {RVWMO Memory Consistency Model, release annotations; Zalasr Load-Acquire/Store-Release extension},
  url    = {https://docs.riscv.org/reference/isa/unpriv/rvwmo.html}
}

@misc{nguyen2026pickle,
  author       = {Nguyen, Hoa and Maidee, Pongstorn and Kaviani, Alireza and Lowe-Power, Jason},
  title        = {Pickle: Precise, Flexible Cross-Core Last-level Cache Data Prefetching for Irregular Memory Accesses},
  year         = {2026},
  eprint       = {2511.19973},
  archivePrefix = {arXiv},
  primaryClass = {cs.AR},
  url          = {https://arxiv.org/abs/2511.19973}
}

@INPROCEEDINGS{yuan2021ioca,
  author={Yuan, Yifan and Alian, Mohammad and Wang, Yipeng and Wang, Ren and Kurakin, Ilia and Tai, Charlie and Kim, Nam Sung},
  booktitle={2021 ACM/IEEE 48th Annual International Symposium on Computer Architecture (ISCA)}, 
  title={Don’t Forget the I/O When Allocating Your LLC}, 
  year={2021},
  volume={},
  number={},
  pages={112-125},
  doi={10.1109/ISCA52012.2021.00018}}

@inproceedings{kim2025a4,
author = {Park, Haneul and Lou, Jiaqi and Lee, Sangjin and Yuan, Yifan and Park, KyoungSoo and Son, Yongseok and Jeong, Ipoom and Kim, Nam Sung},
title = {A4: Microarchitecture-Aware LLC Management for Datacenter Servers with Emerging I/O Devices},
year = {2025},
isbn = {9798400712616},
publisher = {Association for Computing Machinery},
address = {New York, NY, USA},
url = {https://doi.org/10.1145/3695053.3731114},
doi = {10.1145/3695053.3731114},
booktitle = {Proceedings of the 52nd Annual International Symposium on Computer Architecture},
pages = {679–693},
numpages = {15},
location = {
},
series = {ISCA '25}
}

@manual{arm_amba_axi,
  author = {{ARM Limited}},
  title  = {AMBA AXI Protocol Specification},
  year   = {2025},
  note   = {Arm IHI 0022, Issue L, \S A8.4},
  url    = {https://developer.arm.com/documentation/ihi0022/latest/}
}

\end{document}